\documentclass[journal]{IEEEtran}
\usepackage{cite}
\usepackage[pdftex]{graphicx}
\usepackage{amsmath}
\usepackage[caption=false,font=footnotesize]{subfig}
\usepackage{url,times,booktabs, tabularx}
\usepackage[colorlinks=false,pdfborder={0 0 0}]{hyperref}
\usepackage{units}
\usepackage{multirow,array}
\usepackage{balance}
\usepackage{soul}
\usepackage{color}
\usepackage{cleveref}
\usepackage{wrapfig}
\usepackage{booktabs}
\usepackage{balance}
\usepackage[activate]{microtype}
\newcommand{\ie}{i.\,e.,\ }

\begin{document}

%\title{%\fontsize{20}{20}\selectfont
%CAA-Net: Conditional Atrous CNNs with Attention for Visualising Multi-device Acoustic Scenes}
\title{
CAA-Net: Conditional Atrous CNNs with Attention for Explainable Device-robust Acoustic Scene Classification}

\author{Zhao Ren,~\IEEEmembership{Student Member,~IEEE,}
        Qiuqiang Kong,
        Jing Han,~\IEEEmembership{Student Member,~IEEE,}\\
        Mark D. Plumbley,~\IEEEmembership{Fellow,~IEEE,}
        and Bj\"orn W. Schuller,~\IEEEmembership{Fellow,~IEEE}

\thanks{This work was supported by the European Union's Horizon H2020 research and innovation programme under Marie Sk\l{}odowska-Curie grant agreement No.\,766287 (TAPAS), the EPSRC grant EP/N014111/1 ``Making Sense of Sounds'', and the Research Scholarship from the China Scholarship Council (CSC) No.\,201406150082. (Corresponding author: Zhao Ren.)}
\thanks{Z. Ren is with the Chair of Embedded Intelligence for Health Care and Wellbeing, University of Augsburg, Germany (e-mail: zhao.ren@informatik.uni-augsburg.de).}
\thanks{Q. Kong was with the Centre for Vision, Speech and Signal Processing, University of Surrey, UK, when the research has been done, and now is with the ByteDance AI Lab (e-mail: q.kong@surrey.ac.uk).}
\thanks{J. Han was with the Chair of Embedded Intelligence for Health Care and Wellbeing, University of Augsburg, Germany, when the research has been done, and now is with the Department of Computer Science and Technology, University of Cambridge, UK (e-mail: jh2298@cam.ac.uk).}
\thanks{M. Plumbley is with the Centre for Vision, Speech and Signal Processing, University of Surrey, UK (e-mail: m.plumbley@surrey.ac.uk).}
\thanks{B.\ Schuller is with GLAM -- Group on Language, Audio \& Music, Imperial College London, and also with the Chair of Embedded Intelligence for Health Care and Wellbeing, University of Augsburg, Germany (email: schuller@ieee.org).}
}

%\markboth{IEEE Transactions on Multimedia,~Vol.~X, No.~X, November~2020}%

\maketitle

\begin{abstract}
Acoustic Scene Classification (ASC) aims to classify the environment in which the audio signals are recorded. Recently, Convolutional Neural Networks (CNNs) have been successfully applied to ASC. However, the data distributions of the audio signals recorded with multiple devices are different.
There has been little research on the training of robust neural networks on acoustic scene datasets recorded with multiple devices, and on explaining the operation of the internal layers of the neural networks. In this article, we focus on training and explaining device-robust CNNs on multi-device acoustic scene data.
We propose conditional atrous CNNs with attention for multi-device ASC. Our proposed system contains an ASC branch and a device classification branch, both modelled by CNNs. We visualise and analyse the intermediate layers of the atrous CNNs.
A time-frequency attention mechanism is employed to analyse the contribution of each time-frequency bin of the feature maps in the CNNs. On the Detection and Classification of Acoustic Scenes and Events (DCASE) 2018 ASC dataset, recorded with three devices, our proposed model performs significantly better than CNNs trained on single-device data.
%Besides, our experiment demonstrates that the learnt feature maps contain rich information of multi-device acoustic scene data in the time-frequency domain.

\end{abstract}

\begin{IEEEkeywords}
Acoustic Scene Classification, Multi-device Data, Conditional Atrous Convolutional Neural Networks, Attention, Visualisation.
\end{IEEEkeywords}

% \fi
\IEEEpeerreviewmaketitle

\section{Introduction}
\IEEEPARstart{W}{ith} the development of computer audition~\cite{kun2018teaching}, \textit{Acoustic Scene Classification} (ASC) has become a major research field, aiming to automatically recognise acoustic environments~\cite{barchiesi2015acoustic,stowell2015detection}.
The goal of ASC is to identify acoustic scenes in an audio stream, using computational approaches such as signal processing~\cite{qian2017dnn,yang2020learning}, machine learning~\cite{geiger2013large}, and deep learning~\cite{hershey2017cnn,ren2018scalogram}. ASC has been used in various applications, including context-aware services~\cite{perera2014context}, intelligent wearable devices~\cite{harma2003techniques}, and robot navigation systems~\cite{martinson2007robotic}. 

In recent years, deep learning approaches have shown good performance in ASC~\cite{hayashi2016bidirectional,hershey2017cnn}. Deep learning can often learn high-level representations, which yield better performance than those using conventional machine learning methods~\cite{qian2017dnn,amiriparian2017sequence}. Many deep learning structures have been proposed for ASC, including \textit{Convolutional Neural Networks} (CNNs)~\cite{hershey2017cnn} and \textit{Recurrent Neural Networks} (RNNs)~\cite{phan2017audio}. In this article, we will use CNNs due to their ability to learn strong representations from spectrograms~\cite{koluguri2017spectrogram,zhao2019emotion}.
%A spectrogram is a time-frequency representation of an audio wave using the Fourier transformation~\cite{oppenheim1970speech}. 
Specifically, log mel spectrograms have been successfully utilised in ASC~\cite{hershey2017cnn,salamon2017deep}. In this regard, we extract log mel spectrograms as the inputs of the CNNs.

%During the training procedure, data sources can affect classification accuracy. 
%The recording devices usually differ from each other in the microphone characteristics. Therefore, audio data recorded by different devices have various qualities, such as sampling rate, amplitude, and frequency response. The differences in audio qualities can lead to diverse data distributions.
Varying characteristics of recording devices lead to various qualities of audio data, such as sampling rate, amplitude, and frequency response, and data distributions~\cite{primus2019exploiting}.
Audio data is often recorded with distinct devices, such as professional sound recording devices~\cite{brezina2018sound}, and mobile devices~\cite{goyal2019identification}. 
%Recording a large audio dataset with multiple devices can help \zhr{to} train deep neural networks. 
However, it is challenging to optimise the training model on a multi-device audio dataset, due to the different data distributions of mismatched devices~\cite{nguyen2019acoustic}.
% In ASC studies, there are few works focused on processing multi-device audio waves.
%In particular, the task of ASC has the potential to serve in an accessibility service in mobile devices, such as mobile devices contextualization~\cite{battaglino2015acoustic}, and hazard detection~\cite{lee2011acoustic}. Nevertheless, in some cases, the audio data recorded by a single device is not big enough for training deep neural networks. Further, the performance will decrease steeply due to different data distributions, while using a deep learning model trained on single-device audio data to predict the labels of mismatched-device data. 
Supervised domain adaptation by transfer learning can adjust models from the source data to the target data~\cite{kumar2018knowledge}. Yet, it is time-consuming to train two deep models before and after transferring the learnt paramters. 
In this regard, jointly training a single model on multi-device data can reduce the computational complexity compared to training separate models. In previous studies~\cite{Mesaros2018_DCASE,ren2018attention}, researchers trained joint models on multi-device data.
%The shortcoming of joint training is that it cannot learn device-related features
However, joint training cannot analyse the difference of data from mismatched devices, as the deep neural networks learn common parameters for multi-device data~\cite{fourure2017multi}. Inspired by the task of speaker verification~\cite{snyder2016deep,hojo2018dnn}, we propose to condition the CNNs with device information, through feeding the device information into CNNs: we call this conditional training.
%while considering each device as a speaker.
Through conditional training, the CNN model learns specific parameters for the data from each device. To further solve the problem caused by unknown device information in the test set, we show that our proposed multi-task conditional training can also condition the CNNs using predicted device information.

A CNN model generally contains convolutional layers, local pooling layers, and fully connected layers, to achieve a final classification~\cite{ren2018heartsound}. To learn global information from the feature maps using the convolutional kernels, each convolutional layer applies either stride convolutions with strides of length $stride\geq 2$, where strides are the moving steps of convolutional kernels, or local pooling operations. Such CNNs can classify acoustic scenes with state-of-the-art performance~\cite{ren2018attention}.
%However, CNNs cannot \zhr{help us to visualise} \zhr{and} break the black box \zhr{of} CNNs~\cite{ren2018attention}. 
Visualising the operation of the internal layers of CNNs can help us to understand what characteristics are important for classification. For example, the feature maps can be visualised according to the contribution of time-frequency bins to classification. This idea is inspired by image processing tasks, including object localisation~\cite{shi2015bayesian}, and saliency detection~\cite{gao2015database}. %Hence, this article aims to visualise and analyse the high-level representations learnt by CNNs.

Two difficulties emerge when visualising ASC systems. Firstly, the size of the feature maps decreases due to either stride convolution or local pooling layers. Low-resolution feature maps lose time-frequency details compared to the log mel spectrograms. Secondly, the classification result depends on global pooling operations, which summarise the feature maps into fixed dimensional vectors for later classification. 
%Accurate estimation of the contribution of time-frequency bins in feature maps can improve the classification performance. 
Global max and average pooling layers have been widely used in ASC~\cite{ren2018attention}. However, if the value of each bin is viewed as the contribution, the contributions of the bins in feature maps are over- or underestimated in the output of these two kinds of global pooling layers~\cite{ren2018attention}.
To overcome these difficulties, we propose to employ atrous CNNs and an attention mechanism for visualisation. Atrous CNNs can retain the size of the feature maps to be the same as that of the input, by using dilated convolutions~\cite{chen2017rethinking}. Furthermore, an attention mechanism can estimate the contribution of time-frequency bins better than global max or average pooling~\cite{ren2019attention}. Our proposed conditional training framework, which is based on atrous CNNs with attention, is called \textit{Conditional Atrous CNNs with Attention} (CAA-Net).

The remainder of this article is structured as follows. Related work will be introduced in Section~\ref{sec:relatedwork}. In Section~\ref{sec:proposedmethod}, we will describe the proposed CAA-Net, including the framework overview, conditional training, atrous CNNs, and global pooling. The database description, experimental setup and evaluation metrics, experimental results, and the visualisation of the CNNs will then be presented in Section~\ref{sec:experiment}. Finally, the conclusions and future work will be given in Section~\ref{sec:conclusion}.

\section{Related Work}
\label{sec:relatedwork}

Since acoustic scene data from different devices has different qualities and distributions, CNNs trained on different single-device data will differ from each other.
%Therefore, the learnt CNNs differ from each other while each CNN model is trained on single-device data, and the visualisation of CNNs gives various heat maps according to the devices. 
To reduce the computational burden caused by training a separate CNN model for each device, researchers proposed a series of learning strategies. In transfer learning, supervised domain adaptation~\cite{motiian2017unified} trained separate CNN models for the source and target data. In multi-task learning, CNNs shared the parameters of the low-level convolutional layers, and trained the high-level convolutional layers separately~\cite{sindagi2017cnn}. However, more convolutional layers lead to more parameters, so that the training procedure tends to be sub-optimised~\cite{he2015convolutional}. Joint learning with shared context was proposed to process multi-source data using a selective loss function~\cite{fourure2017multi}, but it is challenging to analyse the data difference among multiple devices using joint learning~\cite{fourure2017multi}.
%However, different from multi-task learning, we have only one ASC task based on the multi-device data. 
Conditional CNNs were proposed for multi-pose face recognition~\cite{xiong2015conditional}, where the convolutional kernels of the conditional CNNs were sparsely activated using a conditional activation function in each convolutional layer.
%Similar to conditional CNNs, we propose \zhr{to condition the CNNs by device information}. 
Similar to the idea of conditional CNNs, a deep learning model was conditioned with the speaker information which was fed into the model as an extra input to achieve speech generation~\cite{gibiansky2017deep,van2016wavenet}. 
%Deep neural networks, conditioned by speaker information, were trained on a multi-speaker dataset~\cite{van2016wavenet}. 
Inspired by these approaches~\cite{gibiansky2017deep,van2016wavenet}, we propose to condition CNNs with device information. Furthermore, 
%\zhr{it is difficult to use device information as prior knowledge\zhr{, as} the device information is unknown in practice}. 
multi-task learning was employed for multi-modal emotion recognition, and the predicted difficulty level of the samples during the learning process was fed back into the deep neural networks~\cite{zhang2018dynamic}. Inspired by this work, we propose multi-task conditional training to learn CNNs, predicting acoustic scene classes and device information. %Therefore, our proposed CAA-Net will be trained on multi-device data \zhr{with the device information being either prior knowledge or predicted.}
%while not only the device information is the prior knowledge, but also the device information is unknown.  

%\begin{figure*}[!htb]
%  \centering
%  \includegraphics[width=\linewidth]{figures/fig-framework.pdf}
%  \caption{The framework of our proposed CAA-Net. The log mel spectrogram images are fed into the CNNs, and the device one-hot encoders are expanded from a vector into matrices to be fed into a convolutional layer. Then, the feature maps from the aforementioned two inputs are added to be fed into the next convolutional layer. Finally, a global attention layer follows the convolutional layers for the classification. The dash lines mean the device one-hot encoders are added with the feature maps from part of convolutional layers.}
%  \label{fig:framework}
%\end{figure*}

\begin{figure*}[!htb]
  \centering
  \includegraphics[width=\linewidth]{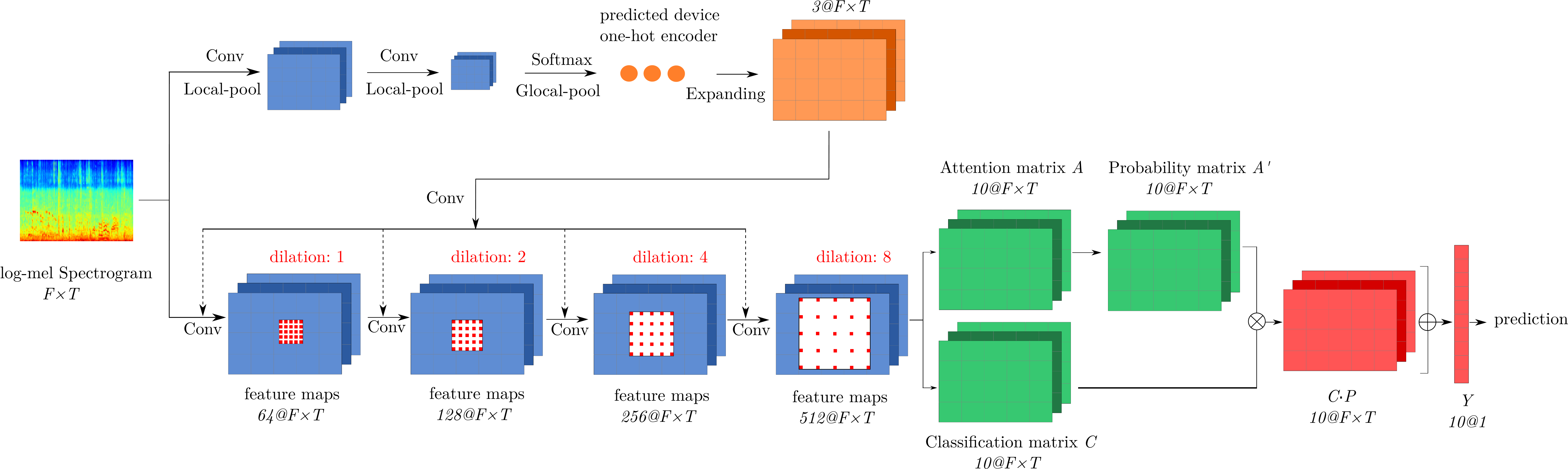}
  \caption{The framework of our proposed multi-task CAA-Net. The log mel spectrograms are fed into two CNN models. The top CNN model predicts the device classes, and the bottom CNN model predicts the acoustic scenes. The predicted device information is transformed into a one-hot encoder, and is then expanded to a three dimensional tensor. Next, the convolutions of the tensors are fed into a convolutional layer of the bottom CNN model. Finally, a global attention layer is applied to the feature maps of the last convolutional layer. The dash lines indicate that the transformed device information is fed into one of the convolutional layers.}
  \label{fig:framework}
\end{figure*}

Many CNN models result in low-resolution feature maps, which cannot describe the time-frequency details~\cite{ren2018attention}. 
%Hence, we try to \zhr{retain} the size of feature maps \zhr{while CNNs go deeper}. 
A basic CNN model without pooling layers can retain the size of the feature maps to be the same as that of the input at each convolutional layer. However, more convolutional operations than those in CNNs with local pooling result in sub-optimised training procedures and lower performances~\cite{ren2018attention}. Some image processing approaches have investigated increasing the size of feature maps at the final convolutional layer to be the same as the input. For example, encoder-decoder CNNs~\cite{vukotic2017one} learnt internal representations by employing encoders and decoders to respectively downsample and upsample the feature maps. \textit{Fully Convolutional Networks} (FCNs)~\cite{ronneberger2015u} utilised deconvolutional layers to upsample the low-resolution feature maps. 
Deconvolutional layers were also applied in the architecture of \textit{Generative Adversarial Networks} (GANs), generating synthetic images which can be used to augment the training data and further improve classification performance~\cite{frid2018gan}.
However, these image processing methods mostly use strongly labelled data, where each pixel of the input image is labelled. The data in ASC is weakly labelled as each audio clip has only one label, rather than each frame being annotated. It is challenging to apply these image processing methods to preserve the same size as the input on the feature maps at the final convolutional layer in ASC~\cite{ren2018attention}.
%\zhr{retain} the size of feature maps in the task of ASC. 
In a separate study of image segmentation~\cite{chen2017rethinking}, atrous CNNs employed dilated convolutions to retain the size of the feature maps to be the same of that of the input at each layer, instead of preserving the size of the feature maps at the final layer only. Inspired by the success of atrous CNNs in image segmentation, we apply atrous CNNs to achieve pixel-wise visualisation of the feature maps for audio classification.

For a classification task, convolutional layers are often followed by global pooling layers~\cite{ren2017dnn}. However, conventional global pooling methods, such as global max and average pooling, cannot estimate the contribution of each time-frequency bin in the feature maps~\cite{ren2017dnn,ren2018attention}. Instead, weighted pooling was proposed to combine different convolutional layers by learning weighted masks, which solves this problem~\cite{hu2017deep,li2018unified}. Similarly, the attention mechanism was proposed for audio and speech-related tasks to estimate the contribution of the feature vectors at each time step~\cite{chorowski2015attention}. Inspired by the attention mechanism idea, we propose to apply a global attention pooling layer to estimate the contribution of each time-frequency bin to the final classification. Finally, our proposed CAA-Net conditions the CNNs with device information, and applies dilated convolutions and an attention mechanism for visualising the intermediate layers of the CNNs.

\section{Proposed Methodology}
\label{sec:proposedmethod}

In this section, we will give an overview of the CAA-Net framework. Then, conditional CNNs will be proposed to adapt the CNNs to handle multi-device data. Finally, we will describe the atrous CNNs and the attention mechanism that we use to visualise the internal layers of the CNNs.

\subsection{Framework Overview}
\label{sec:framework}
The framework of our proposed multi-task CAA-Net is shown in Fig.~\ref{fig:framework}. We extract log mel spectrograms from the multi-device audio waves. Then, the log mel spectrograms are used as input to train our proposed multi-task CAA-Net. CAA-Net contains two branches: one branch aims to predict acoustic scene classes, while the other branch predicts device classes. The first branch contains four convolutional layers and a global attention pooling layer. The four convolutional layers output feature maps with channels of dimensions $64$, $128$, $256$, and $512$. Notably, to preserve the same size of the feature maps as the log mel spectograms, there is no local pooling following each convolutional layer. After the convolutional layers, a global attention pooling layer is applied to learn weight values for each time-frequency bin in the feature maps.

To process multi-device data, the CNN model in the first branch is conditioned by the predicted device information inferred from the CNN model in the second branch. The CNN model in the second branch consists of two convolutional layers with channels of dimensions $64$ and $128$. Each convolutional layer is followed by a local pooling layer. A global max pooling layer is then used to summarise the feature maps to a fixed dimensional vector for classification. The predicted device classes are represented as one-hot encoders, quantifying the integer classes into binary vectors. The one-hot encoders are repeated along the time and frequency axes to form three-dimensional tensors, which are fed into the convolutional layers of the first branch. More details will be introduced in the next subsections.

\subsection{Conditional Training}
\label{sec:conditional_training}
%The distributions of audio data recorded by distinct devices are different. For example, the frequency responses of different mobile devices are diverse. 
We now introduce four training strategies, including single-device training, joint training, teacher forcing conditional training, and multi-task conditional training.

\subsubsection{Single-device Training}
\begin{figure}[!htb]
  \centering
  \includegraphics[width=.95\linewidth]{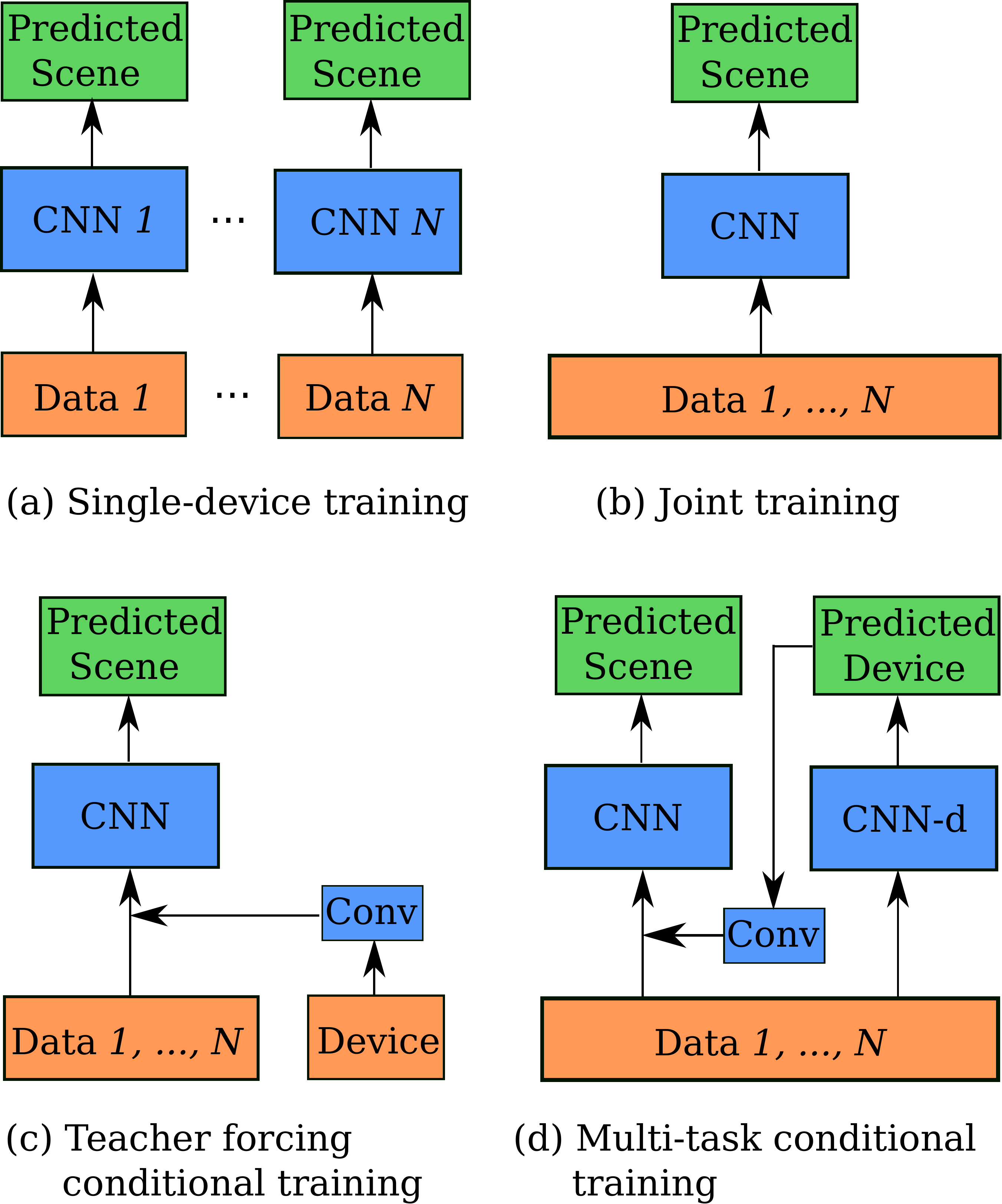}
  \caption{Comparison of the four training strategies. In (d), `CNN-d' means the CNN-device, which aims to predict the device classes.}
  \label{fig:training}
\end{figure}

Single-device training is an approach to train several separate ASC systems on the separate data from each device, as shown in Fig.~\ref{fig:training}(a). With single-device training, several CNN models can be learnt, on each single-device dataset as part of the multi-device dataset~\cite{ren2017dnn}. However, some datasets may be not big enough to train a deep learning model~\cite{valenti2017convolutional}. Additionally, single-device training requires many resources to train $N$ systems for $N$-device data.

\subsubsection{Joint Training}
To overcome the shortcomings of single-device training, we jointly train the audio data from different devices, as shown in Fig.~\ref{fig:training}(b). We assume that data from multiple devices has common features that can be learnt in a CNN model~\cite{fourure2017multi}, and that CNNs can learn complementary representations by jointly training on multi-device data. However, data from different devices share the same network structure and parameters with this approach, so that it is challenging to learn device-related features via joint learning~\cite{fourure2017multi}.

\subsubsection{Teacher Forcing Conditional Training}

To learn device-related representations, the ground truth from a previous time step was fed into the current time step using conditional training in RNNs~\cite{lamb2016professor}. Conditional training has been applied to speech recognition~\cite{van2016wavenet}, and speaker verification~\cite{snyder2017deep}. In our approach, the device information is forwarded to a convolutional layer, and then is fed into the classification branch, as shown in Fig.~\ref{fig:training}(c). In a similar way to a mask, the feature map learnt from the device information gives different weights to each time-frequency bin of the representations in the classification branch, filtering out the time-frequency bins of the feature maps which are not useful. In this regard, the representations computed from the device information are called as `masks' in this article. We call this method teacher forcing conditional training.

The initial mask is generated by expanding the $N$-length device one-hot encoder into a matrix of size $N\times U_w\times U_l$, where $U_w\times U_l$ is the size of the feature maps in the CNN model of the classification branch. The expansion makes it possible to combine the device information with the feature maps. Referring to the linear transformation of the speaker one-hot encoder in a multi-speaker text-to-speech task~\cite{gibiansky2017deep}, we then apply a convolutional layer with a kernel size of $1\times 1$, which is called a 1-by-1 convolution~\cite{ren2017dnn}. Notably, the output of the 1-by-1 convolution has the same number of channels as that of the convolutional layer in the classification branch, with which will be combined. The combined result is given by
\begin{equation}
M_{i+1}=\sigma(W_i*M_i+V_{i}*E_{i}),
\label{eq:condtrain}
\end{equation}
where $M_i$ is the input of the $i$-th convolutional layers, $E_{i}$ is the expanded one-hot encoder, $W_i$ and $V_i$ are the weights of the convolutional procedures, $*$ denotes convolution, and $\sigma$ is the `ReLU' activation function.

\subsubsection{Multi-task Conditional Training}
In teacher forcing conditional training, the device information is required to be fed into the deep neural networks as prior knowledge. However, the device information is sometimes unknown, so we propose to train a model to predict both the acoustic scenes and device classes simultaneously using multi-task learning~\cite{hawthorne2017onsets}, and we call this approach multi-task conditional training. As shown in Fig.~\ref{fig:training}(d), multi-task conditional training learns an additional CNN-device (CNN-d) model to predict the device classes, compared to teacher forcing conditional training. The structure of `CNN-d' model was the same as that introduced in Section~\ref{sec:framework}. The local pooling following each convolutional layer has a kernel size of $2\times 2$ to reduce the size of feature maps and accelerate the training procedure. The predicted device classes, represented as one-hot encoders, are expanded and processed using Eq.~(\ref{eq:condtrain}). In the training procedure, the loss function is computed as the weighted sum of the loss functions in the two CNN models, \ie
\begin{equation}
loss=loss_s + \lambda \times loss_d,
\label{eq:multicondtrain}
\end{equation}
where $loss_s$, and $loss_d$ are the loss values of the CNN model for predicting scenes, and the CNN-d model for predicting devices, respectively. The weight factor $\lambda$ can be use to balance the gradient difference between the two CNN models~\cite{cipolla2018multi}.

\subsection{Atrous Convolutional Neural Networks}
\label{atrouscnn}
To classify the log mel spectrograms representing the audio waveforms, we use atrous CNNs, which can preserve the size of the feature maps to help visualise the internal layers of the CNNs~\cite{ren2018attention}.

\subsubsection{CNNs with local pooling}
\begin{figure}[t]
  \centering
  \includegraphics[width=\linewidth]{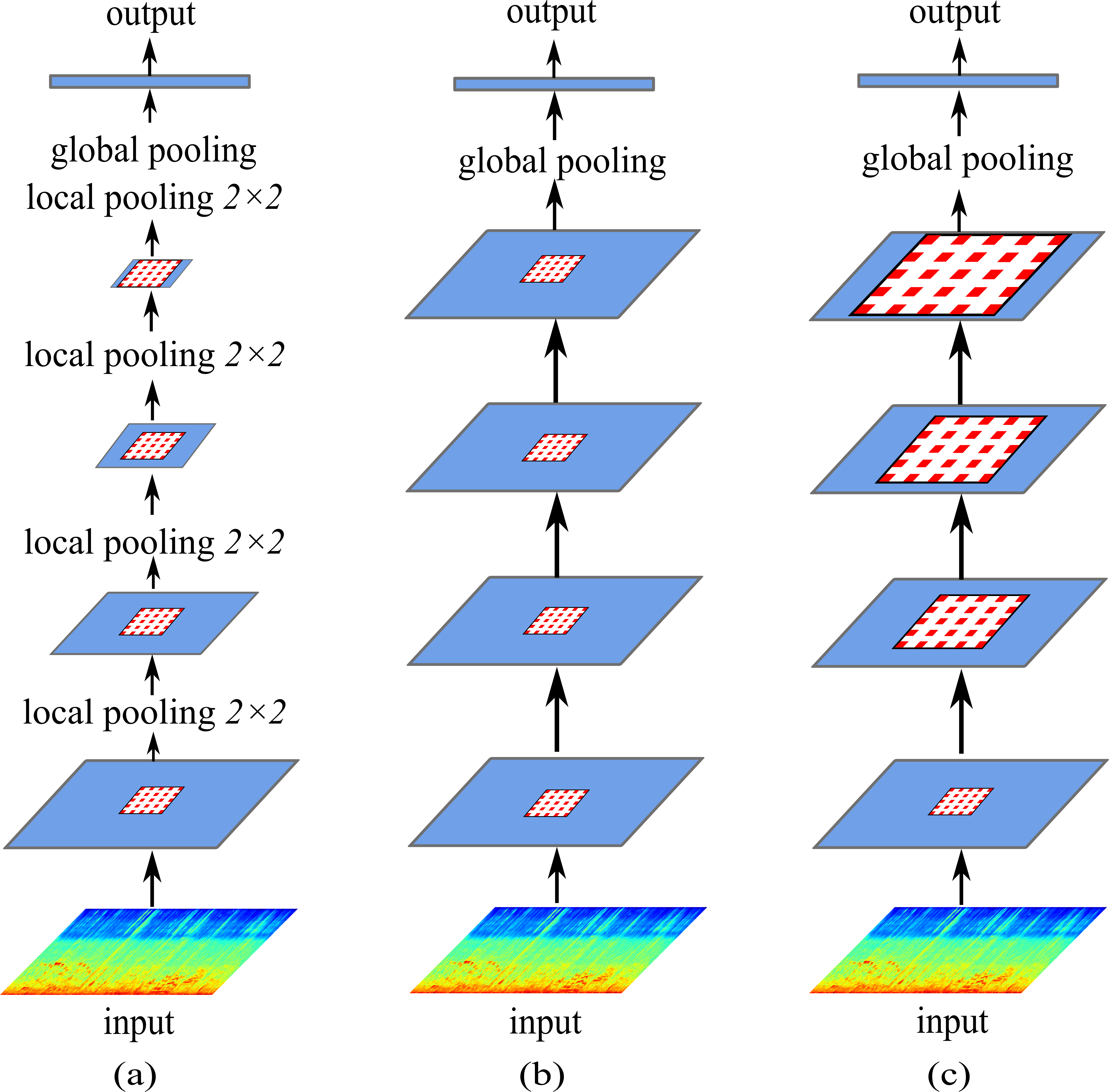}
  \caption{Three CNN architectures: (a) CNNs with local pooling, (b) CNNs without local pooling, and (c) atrous CNNs.}
  \label{fig:cnn}
\end{figure}

In conventional CNNs, the convolutional layers are usually followed by local pooling layers. As shown in Fig.~\ref{fig:cnn}(a), each convolutional layer is followed by a local max pooling layer with a kernel size of $2\times 2$. Local max pooling is able to extract time-frequency shift-invariant features, and can accelerate the training procedure by reducing the size of the feature maps~\cite{phan2018dnn}. However, the local pooling means that the visualisation of the CNNs is limited to low-resolution feature maps.

\subsubsection{CNNs without local pooling}
To solve the problem of low-resolution feature maps in CNNs with local pooling, the local max pooling layers are removed from the CNN model in Fig.~\ref{fig:cnn}(b). This is a basic approach to retain the size of the feature maps, that will lead to a performance reduction~\cite{ren2018attention}. In our previous study~\cite{ren2018attention}, we found that the reason is related to the receptive field,
%The receptive field is the area with a size of (\textit{number of frequency bins, number of time frames}) that can be covered by the kernel during the convolution operation. 
the input area with a size of \textit{number of frequency bins} $\times$ \textit{number of time frames} that can affect a single output pixel in the feature maps. For example, in Fig.~\ref{fig:receptivefield}(a-b), the two receptive fields in the first convolutional layer both have a size of $r\times r$, but in the second convolutional layer, the receptive field of the CNNs with local pooling has a size of $2r\times 2r$, while the receptive field of the CNNs without local pooling still has a size of $r\times r$. Further, at the $l$-th layer, $l\in\{1,..., L\}$ where $L$ is the total number of convolutional layers, the size of the receptive field in the CNNs with local pooling will be $2^{l-1}r \times 2^{l-1}r$, while the CNNs without local pooling will have a fixed receptive field size of $r\times r$ at all layers. Hence, for CNNs with local pooling, the size of the receptive field increases exponentially with the number of layers; but for CNNs without local pooling, it does not change on the number of layers.

\subsubsection{Atrous CNNs}
\begin{figure}[t]
  \centering
  \includegraphics[width=\linewidth]{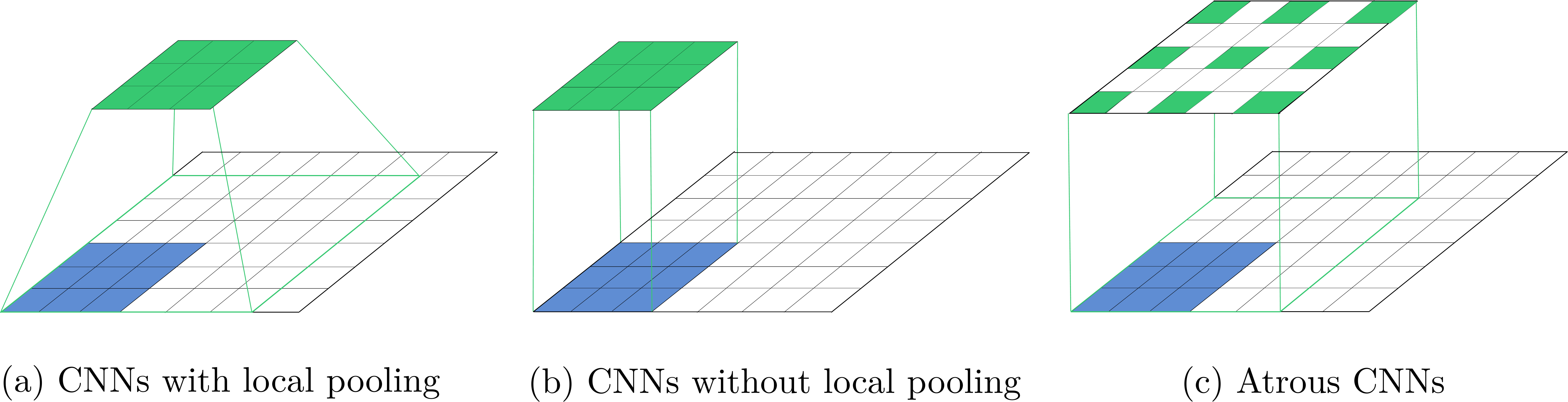}
  \caption{Comparison of the receptive fields in the three CNN architectures. In each subfigure, the white grid at the bottom denotes the feature map in the first convolutional layer, and the blue grid means the convolutional kernel. The top green grid is the kernel in the second convolutional layer.}
  \label{fig:receptivefield}
\end{figure}

To preserve the size of feature maps to be the same as that of the input, we employ atrous CNNs, increasing the size of the receptive field exponentially. In atrous CNNs, the convolutional kernel is replaced by a sparse kernel with holes, called a dilated kernel, the size of which is determined by the dilation rate. The dilation rate in each of the four convolutional layers is set to $1$, $2$, $4$, and $8$, so that the size of the receptive field increases exponentially, the same as in CNNs with local pooling. For example, in Fig.~\ref{fig:receptivefield}(c), the size of the receptive field in the second convolutional layer is $(2r-1)\times (2r-1)$, and $(2^{l-1}r-1)\times (2^{l-1}r-1)$ in the $l$-th layer. The convolutional operations in atrous CNNs are mostly less than those of the CNNs without local pooling.
%, so that the time and space complexities are reduced in atrous CNNs.

\subsection{Global Pooling Mechanism}
\label{globalpooling}
We denote a log mel spectrogram as $S$ of size $F\times T$, where $F$ is number of the frequency bins and $T$ is number of the time frames. The feature maps from the final convolutional layers of the CNNs are denoted as a three-dimensional tensor $M$ of size $H\times P\times Q$, where $H$ denotes the number of channels, and $P\times Q$ is the size of each feature map. The convolutional layers are then followed by a global pooling layer~\cite{he2015spatial}, which summarises each of the $H$ feature maps into a scalar value, which is stacked into an $H$-dimensional vector for classification. Global pooling performs better than flattening the feature maps into a vector, as global pooling can filter out the units from the feature maps which are not useful~\cite{ren2017dnn}. In this subsection, three conventional global pooling methods will be compared, and the proposed attention pooling will be introduced. 

\subsubsection{Global max pooling}
With the assumption that the maximum value can represent the area of each kernel~\cite{amores2013multiple}, global max pooling selects the maximum value from the kernel. Each kernel has the same size as the feature map. Global max pooling can be defined by
\begin{equation}
R_h=\max_{1<q<Q}\max_{1<p<P}M_{hpq},
\end{equation}
\noindent where $R_h$ is the $h$-th value of the vector $R$ which will be fed into a softmax layer for classification.
Global max pooling has achieved success in many applications, such as image processing~\cite{tolias2015particular} and audio event detection~\cite{su2017weakly}. However, global max pooling ignores the contribution of the units with smaller values than the maximum value. The lost information might lead to under-performance of the classification~\cite{ren2017dnn}. 

\subsubsection{Global average pooling}
Different from global max pooling, global average pooling treats every unit as having the same contribution for the classification~\cite{xu2003statistical}. Global average pooling computes the average value of each feature map by
\begin{equation}
R_h=\frac{1}{PQ}\sum_{q=1}^{Q}\sum_{p=1}^{P}M_{hpq}.
\end{equation}
\noindent However, average pooling overestimates the contribution of some units which are not useful for the final classification, and underestimates the contribution of useful units~\cite{ren2017dnn}. Hence, global average pooling is also not suitable to accurately estimate the contribution of each unit.

\subsubsection{Global region of interest pooling}
Global \textit{Region of Interest} (ROI) pooling~\cite{boureau2010theoretical} first splits a feature map into several sub-areas, and then applies global max pooling on these sub-areas. The outputs from these sub-areas are then flattened into a vector for the final classification. Notably, while combining the ROI pooling and the global attention pooling, which will be introduced in the following paragraph, the outputs of the ROI pooling are fed into the attention mechanism instead of flattening, to learn the contribution of each time-frequency bin.
To compare global ROI pooling methods to other pooling methods without considering the effect of hyper-parameters, the size of the sub-areas is set to $16\times 16$.
Therefore, the feature maps after global ROI pooling have the same size as the feature maps from four local max pooling layers with a $2\times 2$ kernel. Global ROI pooling can select more useful units from each feature map. However, like global max and average pooling, it cannot accurately estimate the contribution of each time-frequency bin~\cite{ren2018attention}.

\subsubsection{Global attention pooling}
Compared to the aforementioned pooling methods, the global attention pooling aims to reduce the dimensions of feature maps, by estimating the contribution of each unit~\cite{ren2018attention}. Global attention pooling contains two branches: one is a classification branch, and the other is an attention branch. Global attention pooling can classify the feature maps without a final linear transformation. To achieve the classification in attention pooling, we employ a 1-by-1 convolutional layer in each branch, and the channel number of the output is equal to the class number. Further, in the classification branch, a softmax activation function follows the 1-by-1 convolutional layer for classification. In the attention branch, a sigmoid function is employed. The sigmoid function transforms the values in the feature maps into the interval $[0, 1]$, representing the weight value of each time-frequency bin. In the attention branch, the obtained attention matrix $A$ is normalised using
\begin{equation}
A'_{kpq}=A_{kpq}/\sum_{q=1}^Q\sum_{p=1}^PA_{kpq},
\end{equation}
where $k$ means the $k$-th class, and $A'$ denotes the probability matrix, holding the weight of each element in the H feature maps $C$ from the classification branch. Finally, the prediction $Y$ is computed by
\begin{equation}
Y_k=\sum_{q=1}^Q\sum_{p=1}^PA'_{kpq}\cdot C_{kpq}.
\end{equation}
Notably, the matrices $A'$ and $C$ have the same size, and are multiplied using the element-wise product. In CNNs without local pooling and in atrous CNNs, the feature maps produced by the convolutional layers have the same size as the log mel spectrograms, \ie $P=F$ and $Q=T$.

\section{Experiments and Results}
\label{sec:experiment}
\subsection{Database}

We evaluate our proposed CAA-Net on the open dataset in the ASC task of the IEEE AASP Challenge on Detection and Classification of Acoustic Scenes and Events (DCASE) 2018 and 2019~\cite{Mesaros2018_DCASE}. We test the development set only, as the labels of the test set are not publicly released. 
The DCASE 2018 database was recorded in ten scene environments of six large European cities. The ten acoustic scenes include: \textit{airport}, \textit{shopping mall}, \textit{metro station}, \textit{street pedestrian}, \textit{public square}, \textit{street traffic}, \textit{tram}, \textit{bus}, \textit{metro}, and \textit{park}. Three devices were employed to record the audio waves simultaneously. The main recording device (called device A) comprises a Soundman OKM II Klassik/studio A3, an electret binaural in-ear microphone, and a Zoom F8 audio recorder. The other two recording devices are mobile phones: a Samsung Galaxy S7 (called device B), and an iPhone SE (called device C). During recording, device A was worn in the ears, device B was held in a hand, and device C was worn in a sleeve of the strap of a backpack. The sampling rate of device A was $48$\,kHz, and the sampling rates of device B and C were both $44.1$\,kHz. The original recordings were split into several audio clips with a length of $10$ seconds. Finally, the development set consists of $8,640$ audio files from device A, and $720$ audio files from device B and C in parallel. The ten scene classes are balanced in the whole development set, so that each scene consists of $864$ audio clips from device A, $72$ clips from device B, and $72$ clips from device C. The development set was officially split into a training subset and a test subset by the organisers of DCASE 2018. The training subset contains $6,122$ audio clips from device A, $540$ clips from device B, and $540$ clips from device C. The test subset contains $2,518$ audio clips from device A, $180$ clips from device B, and $180$ clips from device C. 

The DCASE 2019 dataset was recorded in the same ten scene environments as the DCASE 2018 dataset, and extended from six cities to $12$ cities. The three devices and data recording procedures were the same as those in DCASE 2018. Audio clips with a length of $10$ seconds were obtained from the original recordings. The development set contains $16,560$ audio clips in total ($10,265$ audio clips in the training subset, $5,265$ clips in the test subset, and $1,030$ clips from Milan). The training subset consists of $9,185$ audio clips from device A, and $540$ clips from devices B and C in parallel. The test subset contains $4,185$ audio clips from device A, and $540$ clips from devices B and C in parallel. As the data from Milan was not split into the training or test subsets due to data balance issue in the challenge, only the official training and test subsets are used in our experiments.

\subsection{Experimental Setup and Evaluation Metrics}
The audio clips are first resampled to $44.1$ kHz. Then, the log mel spectrograms are extracted from the audio clips using a Hamming window of $2,048$ samples length, an overlap of $672$, and $64$ mel bins. Hence, log mel spectrograms with a size of $64\times 320$ are obtained, where $320$ is the number of time frames. The log mel spectrograms are then fed into the CNN model for classification. The CNN model is optimised using an Adam optimiser with a learning rate of $0.001$. During training, the learning rate is reduced by a factor of $0.9$ at each $200$-th iteration. Finally, the training procedure is stopped at the $15,000$-th iteration. Notably, in the multi-task conditional training, the $\lambda$ value in Eq.~(\ref{eq:multicondtrain}) is set to $1$ and $0.0001$ empirically before and after the accuracy of the `CNN-d' on the test subset achieves $98\,\%$, respectively. 
The setting of $\lambda$ aims to obtain a robust `CNN-d' before training the CNN model for the ASC task.
All of these training hyperparameters are chosen empirically. The source code of our work is publicly released\footnote{\url{https://github.com/EIHW/CAANet_DCASE_ASC}}.

For the evaluation of the performance, we use the unweighted average of class-wise accuracies, which are the number of correctly classified clips divided by the total number of clips in each class. Accuracy is the official DCASE evaluation metrics for ASC.

\begin{figure*}[t]
\centering
\includegraphics[width=\linewidth]{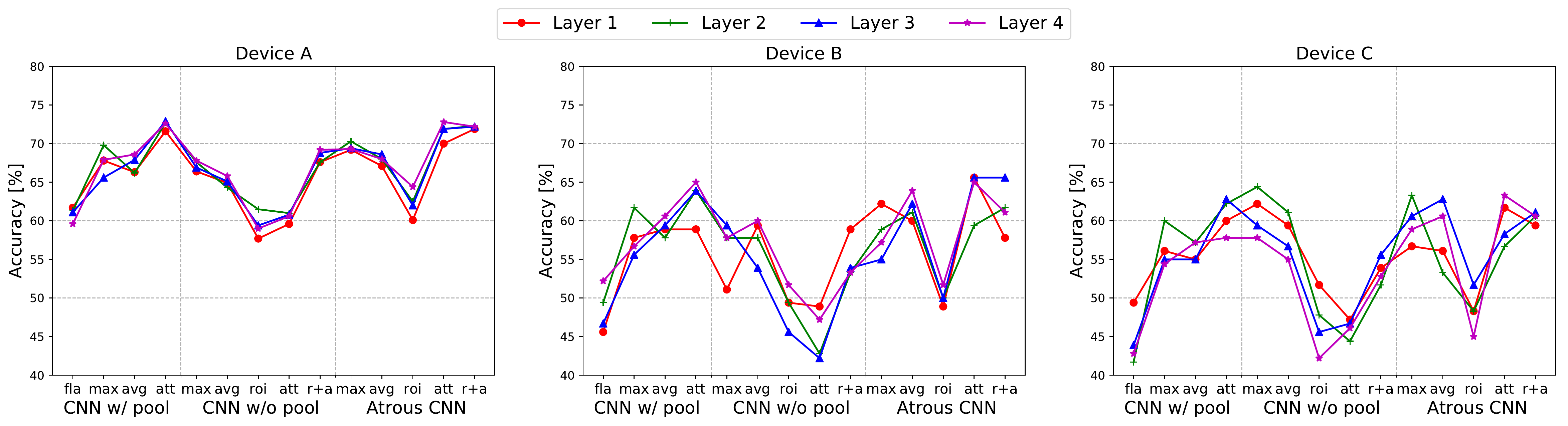}
\caption{Performance (accuracy) comparison of the CNN models evaluated on the DCASE 2018 dataset, while teaching forcing conditional training works at different convolutional layers. The three CNN topologies contain CNNs with local pooling, CNNs without local pooling, and atrous CNNs. The CNNs are followed by flattening (`fla') and five global pooling models, including max, average (`avg'), ROI, attention (`att'), and the combination of ROI and attention (`r+a'). The performance is evaluated on the data from the three devices \textit{A}, \textit{B}, and \textit{C}.}
\label{fig:result1}
\end{figure*}

\begin{table}[!htp]
  \caption{The average performance (accuracy) comparison of the CNN topologies evaluated on the data from the three devices (\textit{A}, \textit{B}, and \textit{C}) available in the DCASE 2018 dataset, while the teaching forcing conditional training works at different convolutional layers. The three types of CNN topologies contain CNNs with local pooling, CNNs without local pooling, and atrous CNNs. The CNNs are followed by flattening and five global pooling models, including max, average (`avg'), ROI, attention (`att'), and the combination of ROI and attention (`roi+att'). The best result chosen from four layers in each CNN model is highlighted.}
  \vspace{5pt}
  \label{tab:result1}
  \centering
  \begin{tabular}{l l |p{.95cm}p{.95cm}p{.95cm}p{.95cm}}
    \toprule
    %\textbf{\textsc{Accuracy [\%]}} & & \multicolumn{1}{c|}{\textbf{\textsc{Layer 1}}} & \multicolumn{1}{c|}{\textbf{\textsc{Layer 2}}} & \multicolumn{1}{c|}{\textbf{\textsc{Layer 3}}} &
    %\multicolumn{1}{c}{\textbf{\textsc{Layer 4}}}\\
%	\cmidrule(lr){1-18} 
%    Network & Pooling& $A$ & $B$ & $C$ & $Avg$ & $A$ & $B$ & $C$ & $Avg$& $A$ & $B$ & $C$ &$Avg$& $A$ & $B$ & $C$ &$Avg$\\
     \textbf{Accuracy [\%]} && \textbf{Layer 1} & \textbf{Layer 2} & \textbf{Layer 3} & \textbf{Layer 4} \\
%    \midrule
%    Baseline & & .597& .589& .451& .462\\
	\midrule
	CNN w/ pool & flatten   &\textbf{52.2} &50.8 &50.6 &51.5  \\
    CNN w/ pool & max       &60.6 &\textbf{63.8} &58.7 &59.7  \\
    CNN w/ pool & avg       &60.1 &60.4 &60.8 &\textbf{62.1}  \\
    CNN w/ pool & att       &63.5 &66.3 &\textbf{66.5} &65.1 \\
    \midrule
    CNN w/o pool& max       &59.9 &\textbf{63.3} &61.9 &61.1\\
    CNN w/o pool& avg       &\textbf{61.2} &61.1 &58.6 &60.3\\
    CNN w/o pool& roi       &\textbf{52.9} &52.9 &50.2 &51.0 \\
    CNN w/o pool& att       &\textbf{51.9} &49.4 &49.9 &51.3  \\
    CNN w/o pool& roi+att   &\textbf{60.1} &57.5 &59.4 &58.4  \\
    \midrule
    Atrous CNN & max        &62.7 &\textbf{64.2} &61.7 &61.8 \\
    Atrous CNN & avg        &61.1 &60.8 &\textbf{64.5} &64.2 \\
    Atrous CNN & roi        &52.4 &53.6 &\textbf{54.6} &53.7 \\
    Atrous CNN & att        &65.8 &62.7 &65.3 &\textbf{67.0} \\
    Atrous CNN & roi+att    &63.0 &64.9 &\textbf{66.3} &64.6 \\
	\bottomrule
\end{tabular}
\end{table}

\begin{table*}[!htb]
  \caption{Performance (accuracy) comparison of the CNN topologies assessed on the DCASE 2018 dataset using four training strategies, including single-device training, joint training, teacher forcing conditional training, and multi-task conditional training. The pooling methods are the same as the methods in Table~\ref{tab:result1}. The performance is evaluated on the data from the three devices \textit{A}, \textit{B}, and \textit{C}. The best result among the four training strategies for each CNN model is highlighted.}
  \vspace{5pt}
  \label{tab:result2}
  \centering
  \begin{tabular}{l l |p{.5cm}p{.5cm}p{.5cm}p{.5cm}| p{.5cm}p{.5cm}p{.5cm} p{.5cm}|p{.5cm}p{.5cm} p{.5cm}p{.5cm}|p{.5cm}p{.5cm} p{.5cm}p{.5cm}}
    \toprule
    \multirow{2}{*}{\textbf{\textsc{Accuracy [\%]}}} & & \multicolumn{4}{c|}{\multirow{2}{*}{\textbf{\textsc{Single-device Training}}}} & \multicolumn{4}{c|}{\multirow{2}{*}{\textbf{\textsc{Joint Training~\cite{ren2019attention}}}}} & \multicolumn{8}{c}{\textbf{\textsc{Conditional Training}}} \\
    %\cline{11-18}
    \cmidrule(lr){11-18}  
    &&&&&&&&&& \multicolumn{4}{c|}{\textbf{\textsc{Teacher Forcing}}}  & \multicolumn{4}{c}{\textbf{\textsc{Multi-task}}} \\
	\cmidrule(lr){1-18}  
    Network & Pooling& $A$ & $B$ & $C$  & $Avg$ & $A$ & $B$ & $C$ & $Avg$ & $A$ & $B$ & $C$& $Avg$ & $A$ & $B$ & $C$& $Avg$ \\
%    \midrule
%    Baseline & & .597& .589& .451& .462\\
	\midrule
	CNN w/ pool & flatten &61.1 &49.4&40.6 &50.4 &61.6 &49.4 &46.7 &52.6 &61.7 &45.6 &49.4 &52.2 &\textbf{63.5} &\textbf{52.8} &\textbf{50.6} &\textbf{55.6}\\
    CNN w/ pool  & max    &70.0 &57.8 &56.1 &61.3 &69.8 &57.2 &57.8 &61.6 &\textbf{69.8} &\textbf{61.7} &\textbf{60.0} &\textbf{63.8} &68.0 &58.3 &57.8 &61.4\\
    CNN w/ pool  & avg    &63.6 &50.0 &62.8 &58.8 &65.8 &57.2 &57.8 &60.3 &\textbf{68.6} &\textbf{60.6} &\textbf{57.2} &\textbf{62.1} &65.4 &60.0 &58.3 &59.7\\
    CNN w/ pool  & att    &73.1 &51.1 &58.3 &60.8 &72.6 &62.2 &56.1 &63.6 &\textbf{72.9} &\textbf{63.9} &\textbf{62.8} &\textbf{66.5} &72.6 &60.6 &59.4 &64.2 \\
    \midrule
    CNN w/o pool& max     &67.8 &53.9 &56.1 &59.3 &61.9 &46.7 &52.2 &53.6 &\textbf{67.6} &\textbf{57.8} &\textbf{64.4} &\textbf{63.3} &65.4 &56.7 &57.2 &59.8\\
    CNN w/o pool& avg     &61.9 &61.7 &42.8 &55.5 &59.1 &54.4 &50.0 &54.5 &\textbf{64.9} &\textbf{59.4} &\textbf{59.4} &\textbf{61.2} &66.1 &61.7 &55.4 & 61.1\\
    CNN w/o pool& roi     &60.4 &43.3 &43.3 &49.0 &61.7 &50.6 &43.9 &52.1 &57.7 &49.4 &51.7 &52.9 &\textbf{60.7} &\textbf{50.0} &\textbf{48.9} & \textbf{53.2} \\
    CNN w/o pool& att     &\textbf{62.7} &\textbf{56.7} &\textbf{60.0} &\textbf{59.8} &59.6 &45.0 &43.3 &49.3 &59.6 &48.9 &47.2 &51.9 &59.2 &46.1 &49.4 &51.6\\
    CNN w/o pool& roi+att &\textbf{67.8} &\textbf{61.1} &\textbf{60.0} &\textbf{63.0} &69.2 &56.1 &50.6 &58.6 &67.6 &58.9 &53.9 &60.1 &67.5 &58.3 &53.3 &59.7\\
    \midrule
    Atrous CNN & max      &66.2 &60.6 &42.8 &56.5 &69.7 &60.0 &59.4 &63.0 &\textbf{70.3} &\textbf{58.9} &\textbf{63.3} &\textbf{64.2} &67.0 &61.1 & 58.9 &62.3 \\
    Atrous CNN & avg      &69.5 &61.1 &55.6 &62.1 &67.2 &62.8 &60.0 &63.3 &\textbf{68.6} &\textbf{62.2} &\textbf{62.8} &\textbf{64.5} &66.9 &56.1 &55.0 & 59.3 \\
    Atrous CNN & roi      &61.2 &45.6 &40.6 &49.1 &62.6 &48.3 &43.9 &51.6 &\textbf{62.0} &\textbf{50.0} &\textbf{51.7} &\textbf{54.6}&63.5 &47.2 &45.0 &51.9 \\
    Atrous CNN & att      &70.7 &61.1 &63.3 &65.0 &73.2 &64.4 &62.2 &66.6 &72.8 &65.0 &63.3 &67.0 &\textbf{72.4} &\textbf{68.9} &\textbf{62.8} &\textbf{68.0} \\
    Atrous CNN & roi+att  &72.7 &58.9 &49.4 &60.3 &72.2 &57.2 &56.7 &62.0 &\textbf{72.2} &\textbf{65.6} &\textbf{61.1} &\textbf{66.3} &72.2 &56.7 &62.2 &63.7 \\
	\bottomrule
\end{tabular}
\end{table*}

\begin{table*}[!htb]

  \caption{Performance (accuracy) comparison of the CNN topologies assessed on the DCASE 2019 dataset in the four training strategies, including single-device training, joint training, teacher forcing conditional training, and multi-task conditional training. The pooling methods are the same as the methods in Table~\ref{tab:result1}. The performance is evaluated on the data from the three devices \textit{A}, \textit{B}, and \textit{C}. The best result among the four training strategies for each CNN model is highlighted.}
  \vspace{5pt}
  \label{tab:result3}
  \centering
  \begin{tabular}{l l |p{.5cm}p{.5cm}p{.5cm}p{.5cm}| p{.5cm}p{.5cm}p{.5cm} p{.5cm}|p{.5cm}p{.5cm} p{.5cm}p{.5cm}|p{.5cm}p{.5cm} p{.5cm}p{.5cm}}
    \toprule
    \multirow{2}{*}{\textbf{\textsc{Accuracy [\%]}}} & & \multicolumn{4}{c|}{\multirow{2}{*}{\textbf{\textsc{Single-device Training}}}} & \multicolumn{4}{c|}{\multirow{2}{*}{\textbf{\textsc{Joint Training}}}} & \multicolumn{8}{c}{\textbf{\textsc{Conditional Training}}} \\
    %\cline{11-18}
    \cmidrule(lr){11-18}  
    &&&&&&&&&& \multicolumn{4}{c|}{\textbf{\textsc{Teacher Forcing}}}  & \multicolumn{4}{c}{\textbf{\textsc{Multi-task}}} \\
	\cmidrule(lr){1-18}  
    Network & Pooling& $A$ & $B$ & $C$  & $Avg$ & $A$ & $B$ & $C$ & $Avg$ & $A$ & $B$ & $C$& $Avg$ & $A$ & $B$ & $C$& $Avg$ \\
	\midrule
    CNN w/ pool  & att    &67.2 &47.0 &47.0 &53.7 &69.9 &44.1 &49.8 &54.6 &70.9 &46.7 &48.1 &55.2 &\textbf{70.9} &\textbf{46.1} &\textbf{49.1} &\textbf{55.4} \\
    CNN w/o pool& att     &60.9 &48.9 &48.3 &52.7 &63.5 &36.9 &36.5 &45.6 &\textbf{65.0} &\textbf{50.6} &\textbf{56.3} &\textbf{57.3} &63.3 &43.3 &46.3 &51.0 \\
    Atrous CNN & att      &72.2 &45.4 &50.2 &55.9 &71.4 &44.3 &50.9 &55.5 &72.0 &43.7 &51.1 &55.6 &\textbf{71.3} &\textbf{44.4} &\textbf{52.6} &\textbf{56.1} \\
	\bottomrule
\end{tabular}

\end{table*}

\subsection{Results and Discussion}
We tested the performance of the three CNN models introduced in Section~\ref{atrouscnn} on the DCASE 2018 dataset, while conditioning the device information at four convolutional layers using teacher forcing conditional training. A set of global pooling methods introduced in Section~\ref{globalpooling} were applied for each CNN model. The performances of the CNN models on the data from the three devices are compared in Fig.~\ref{fig:result1}. In Fig.~\ref{fig:result1}, the performances of the CNN models on device A are better than those on devices B and C, which might be caused by the data imbalance among the data from devices A, B, and C, and the potential existence of more noise in the data from the mobile devices (B and C). While comparing the performances of all CNN models in each device, the atrous CNNs mostly perform better than CNNs with or without local pooling.
%\st{, as atrous CNNs can adapt the receptive field increasing exponentially with convolutional layers}.
The results from conditioning the device information at different convolutional layers are similar on each CNN model. 

To better compare the performance of all CNN models and choose the best convolutional layer to achieve multi-task conditional training by each CNN model, the average performances on the data of the three devices are calculated in Table~\ref{tab:result1}. As shown in Table~\ref{tab:result1}, in the CNNs with local pooling and atrous CNNs, we can obtain better results from global attention pooling than from other global pooling methods. This indicates that attention pooling is suitable to estimate the contribution of each time-frequency bin in the feature maps to the final classification. Further, conditioning the device information at different layers can affect the classification accuracy. Especially, in CNNs without local pooling, most models perform best when the transformed device information is fed into the first convolutional layer. In atrous CNNs, the models in which the device information was conditioned at the third convolutional layer produce better results than those conditioned in other layers. 
%Each CNN model can lead to various results while the device information was conditioned at different layers. 
We select the best results from the four conditioned layers for later comparison of the four training strategies in Section~\ref{sec:conditional_training}.

In Table~\ref{tab:result2}, teacher forcing conditional training and multi-task conditional training are compared to the other two training strategies, including single-device training and joint training. 
%The results of teacher forcing conditional training were selected from the best results during conditioning the device number in four convolutional layers. 
The results of multi-task conditional training were obtained while conditioning in the same convolutional layers as teacher forcing conditional training. The two conditional training strategies perform the best in most CNN models, except global attention and global ROI, and attention pooling in CNNs without local pooling. This may be caused by the large number of 
%parameters
convolutional operations
in those two global pooling methods in CNNs without local pooling. Teacher forcing conditional training performs slightly better than multi-task conditional training. This is reasonable, as the true device information is fed into the CNNs during teacher forcing conditional training. For devices B and C, the performance of joint training is comparable with that of single-device training. This indicates that joint training can learn complementary features for each single-device data. In contrast, conditional training can learn not only complementary features, but also device-related representations. Hence, in our results, conditional training can effectively improve the performance on devices B and C. Finally, our best result of $68.0\,\%$ on average was obtained by multi-task conditional training. It is significantly higher than the result of $65.0\,\%$ from the atrous CNNs with attention on a single device (in a one-tailed z-test, $p<0.01$), and the result of $49.0\,\%$ from the CNNs without local pooling with ROI pooling on a single device (in a one-tailed z-test, $p<0.001$).

\begin{figure*}[t]
\centering
\begin{minipage}[b]{.32\linewidth}
  \centering
  \centerline{\includegraphics[width=\linewidth]{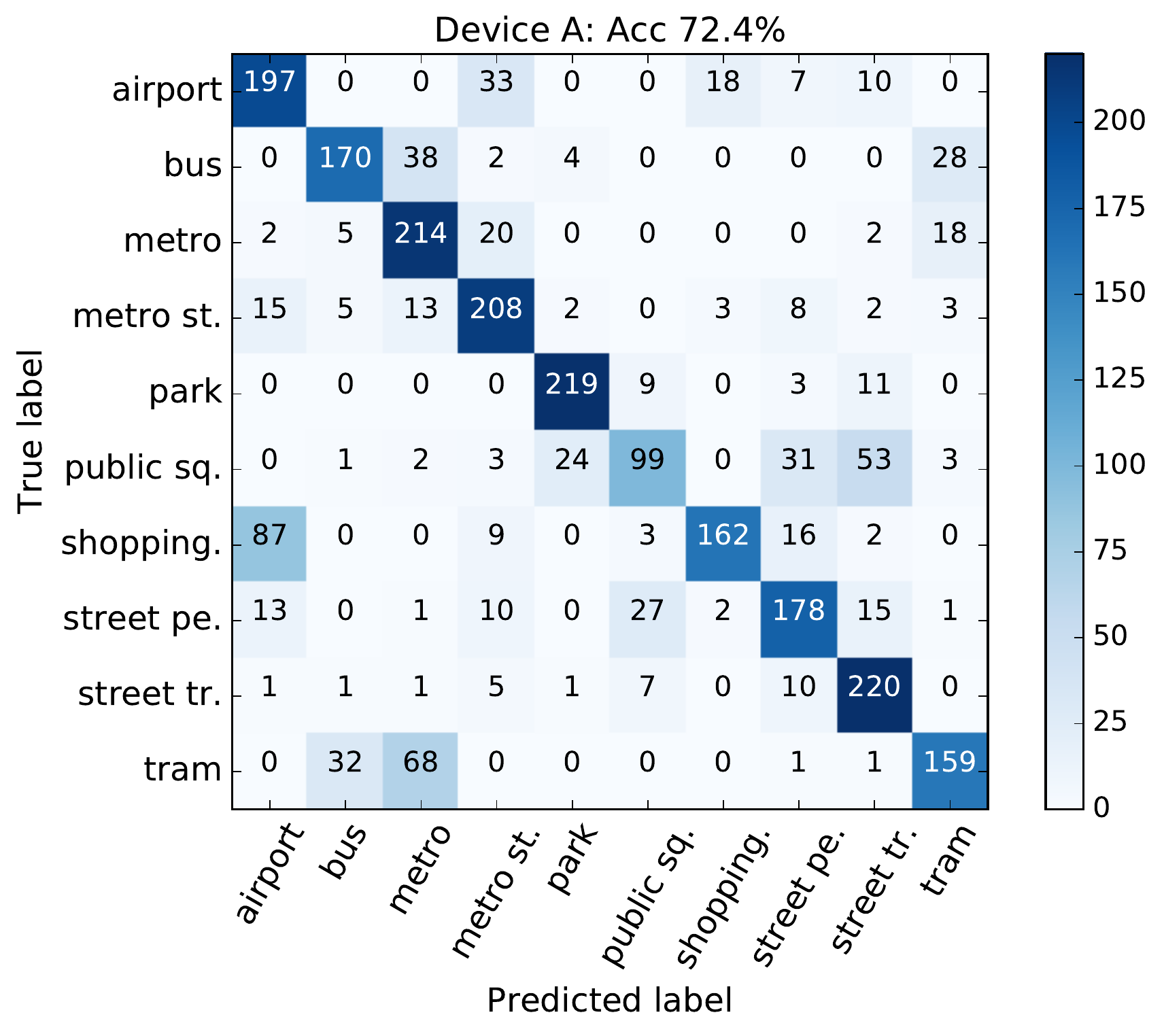}}
%  \centerline{(a)}\medskip
\end{minipage}
\hfill
\begin{minipage}[b]{.32\linewidth}
  \centering
  \centerline{\includegraphics[width=\linewidth]{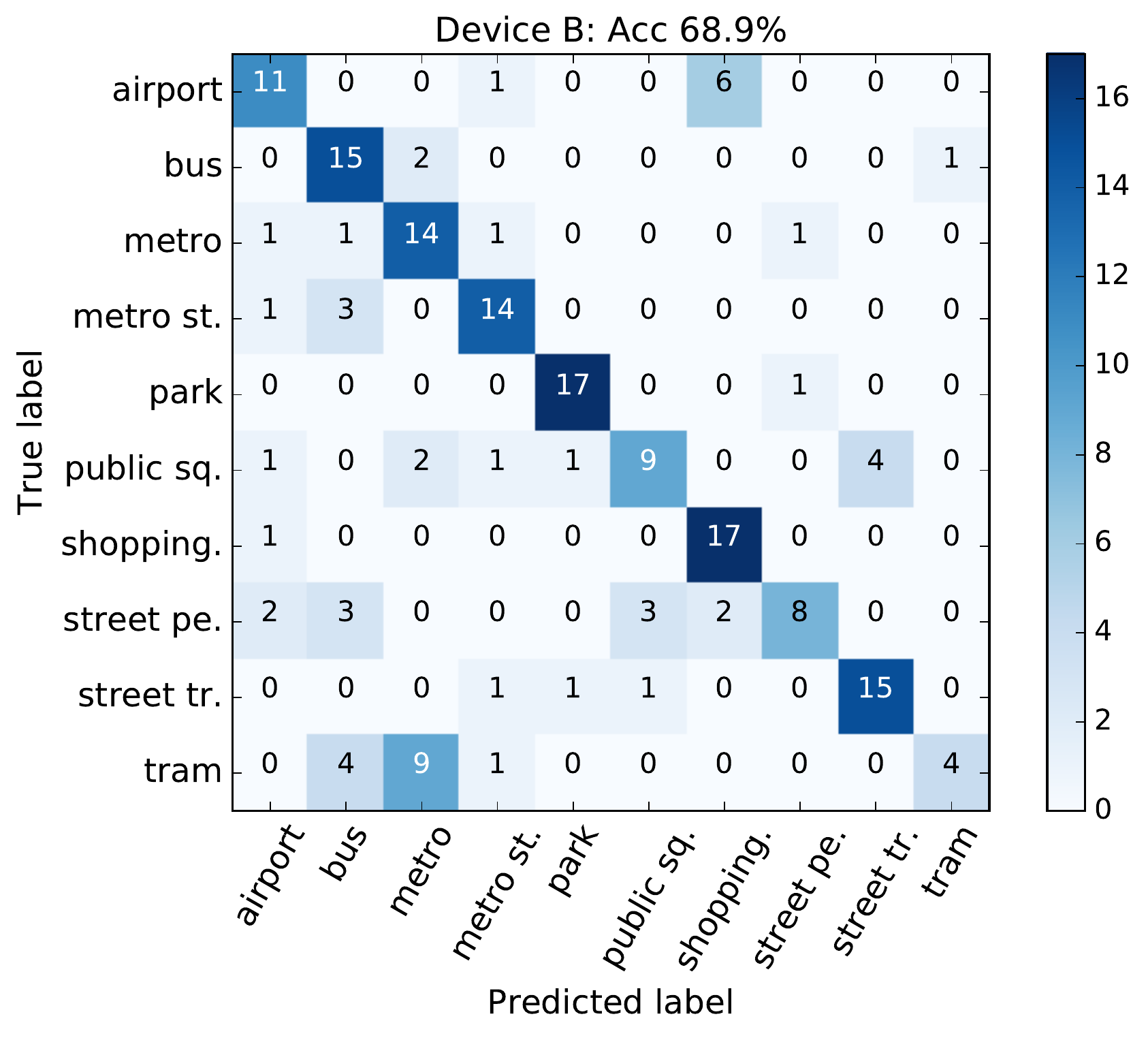}}
%  \centerline{(b)}\medskip
\end{minipage}
\hfill
\begin{minipage}[b]{.32\linewidth}
  \centering
  \centerline{\includegraphics[width=\linewidth]{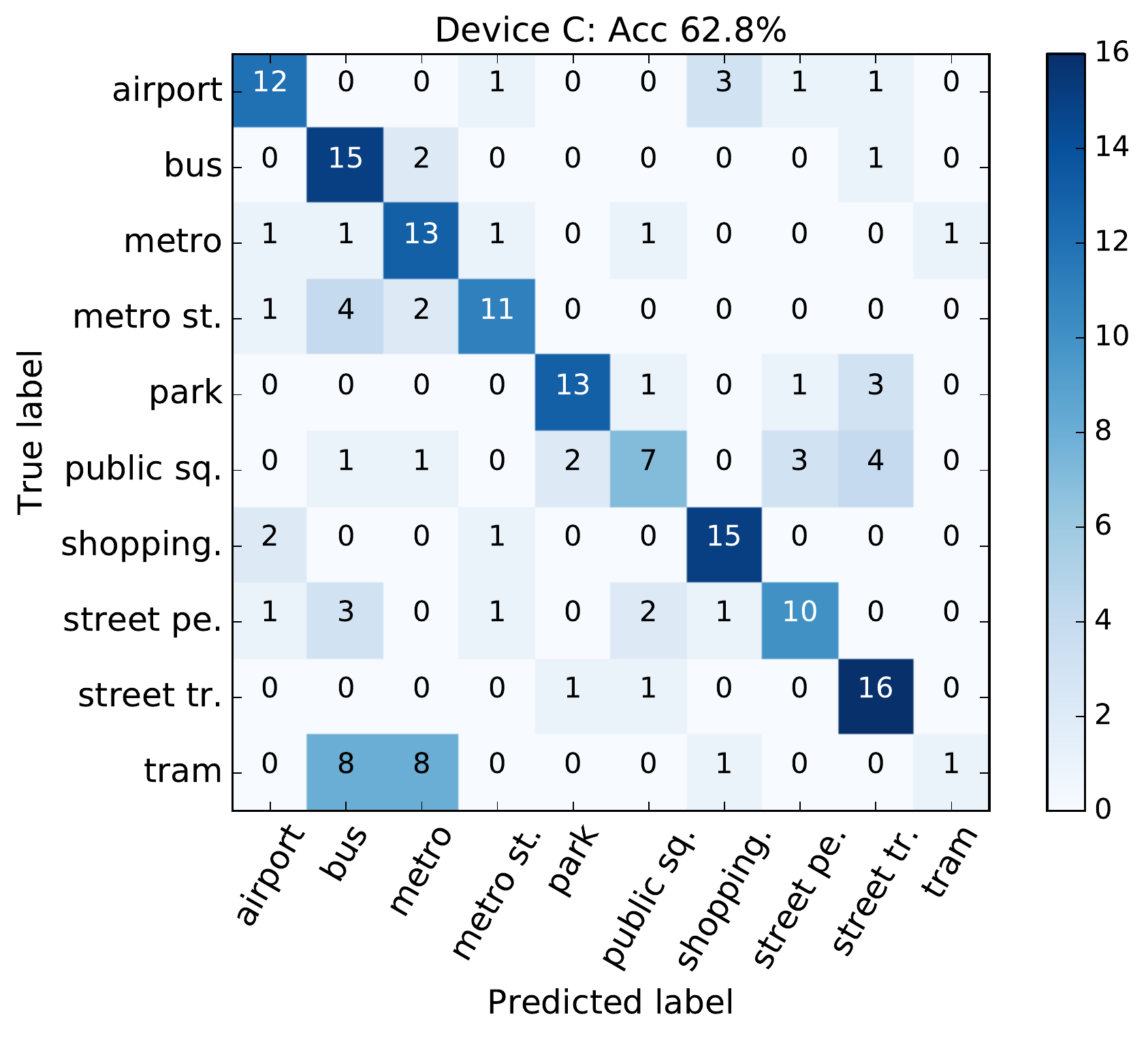}}
%  \centerline{(b)}\medskip
\end{minipage}
\caption{Confusion matrices of the results on the data from devices \textit{A}, \textit{B}, and \textit{C} computed by the proposed multi-task CAA-Net model, which achieves the best result on the DCASE 2018 dataset.}
\label{fig:confmat}
\end{figure*}

Finally, the confusion matrices of the best results obtained by our proposed multi-task CAA-Net are shown in Fig.~\ref{fig:confmat}. The proposed model performs the best on the data from device A, perhaps because the database contains more data for device A than B and C, and the data recorded with device A has a higher quality than the other data. The CAA-Net performs well in some classes, such as \textit{metro}, \textit{park}, and \textit{street traffic}, but for not well for other classes, including \textit{public square}, and \textit{street pedestrian}. The reason may be that the audio recordings in these two environments contain more noise. Furthermore, some classes of scenes are easy to be predicted as other similar scenes. For instance, the scene class \textit{tram} is mostly predicted into the classes of \textit{bus} and \textit{metro}, especially on the data from devices B and C, perhaps because the audio recordings from these three scenes are all traffic sounds. This suggests that besides the types of devices, the type of acoustic scene might also be a factor to be considered in this classification task.

To verify the robustness of our proposed approach, the three CNN models with attention were trained on the DCASE 2019 dataset, as shown in Table~\ref{tab:result3}. For conditional training, the CNN layers corresponding to those highlighted in Table~\ref{tab:result1} were conditioned in the CNN models to predict the acoustic scenes. We can see that, in a similar way to the work on the DCASE 2018 dataset, conditional training achieves the best results in the three CNN models assessed on the DCASE 2019 dataset. The consistent results indicate that our proposed approach is more effective and robust than single-device training and joint training on multiple datasets.

\begin{figure*}[!htbp]
    \centering
  \begin{minipage}[b]{\linewidth}
    \includegraphics[width=.99\linewidth]{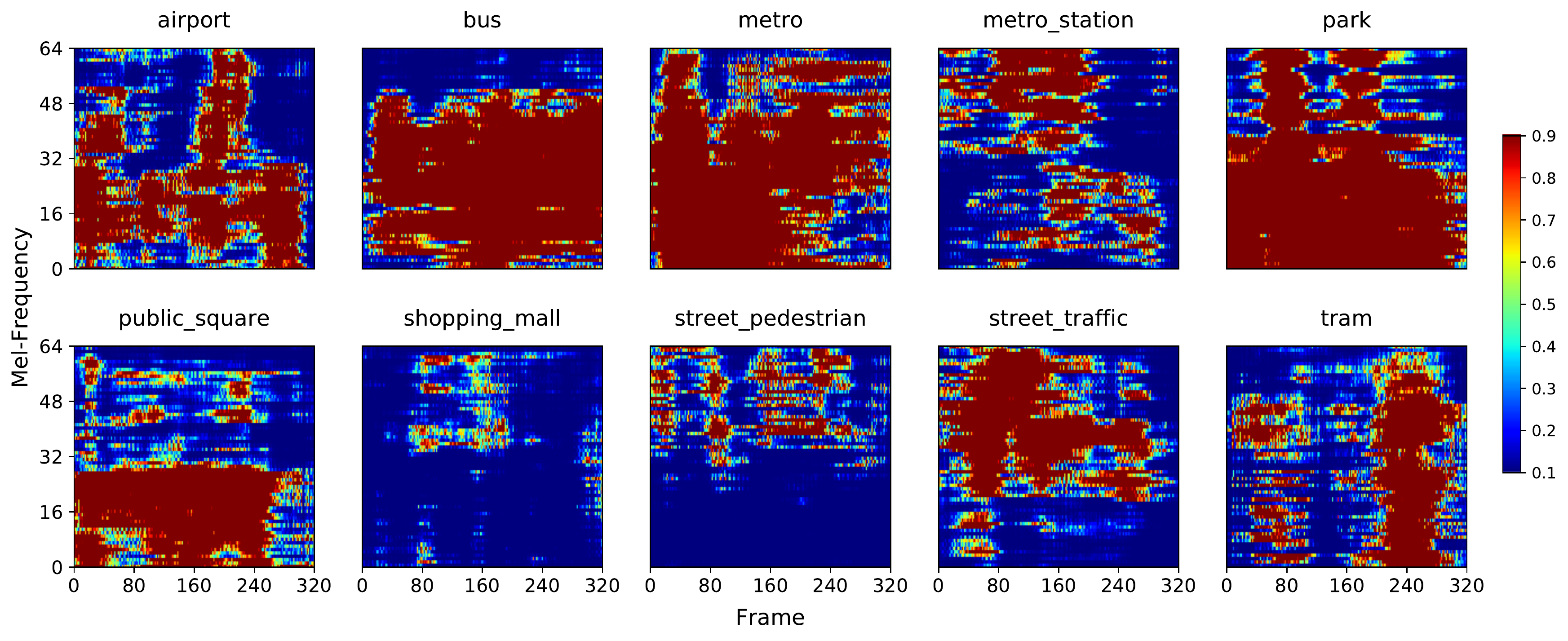}
    \centerline{(a) Heat maps of CNNs on data from device A}
  \end{minipage}
  \begin{minipage}[b]{\linewidth}
    \includegraphics[width=.99\linewidth]{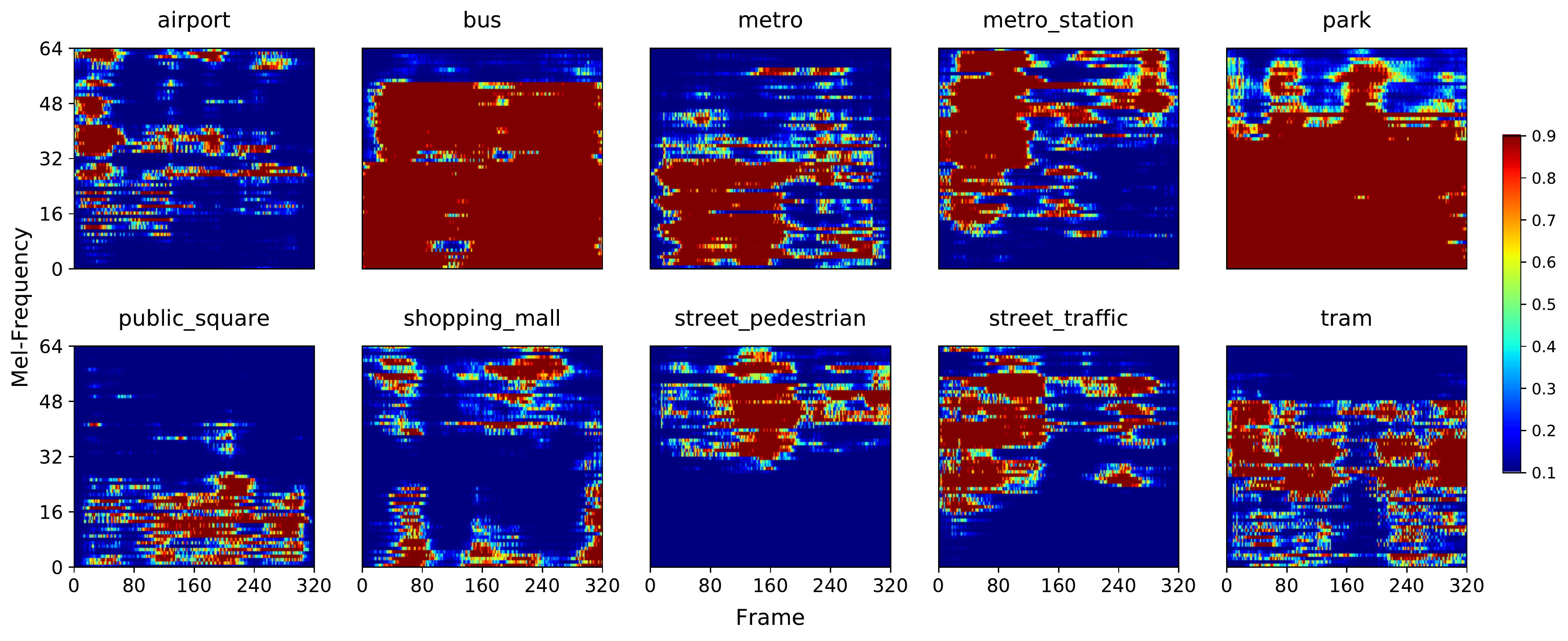}
    \centerline{(b) Heat maps of CNNs on data from device B}
  \end{minipage}
  \begin{minipage}[b]{\linewidth}
    \includegraphics[width=.99\linewidth]{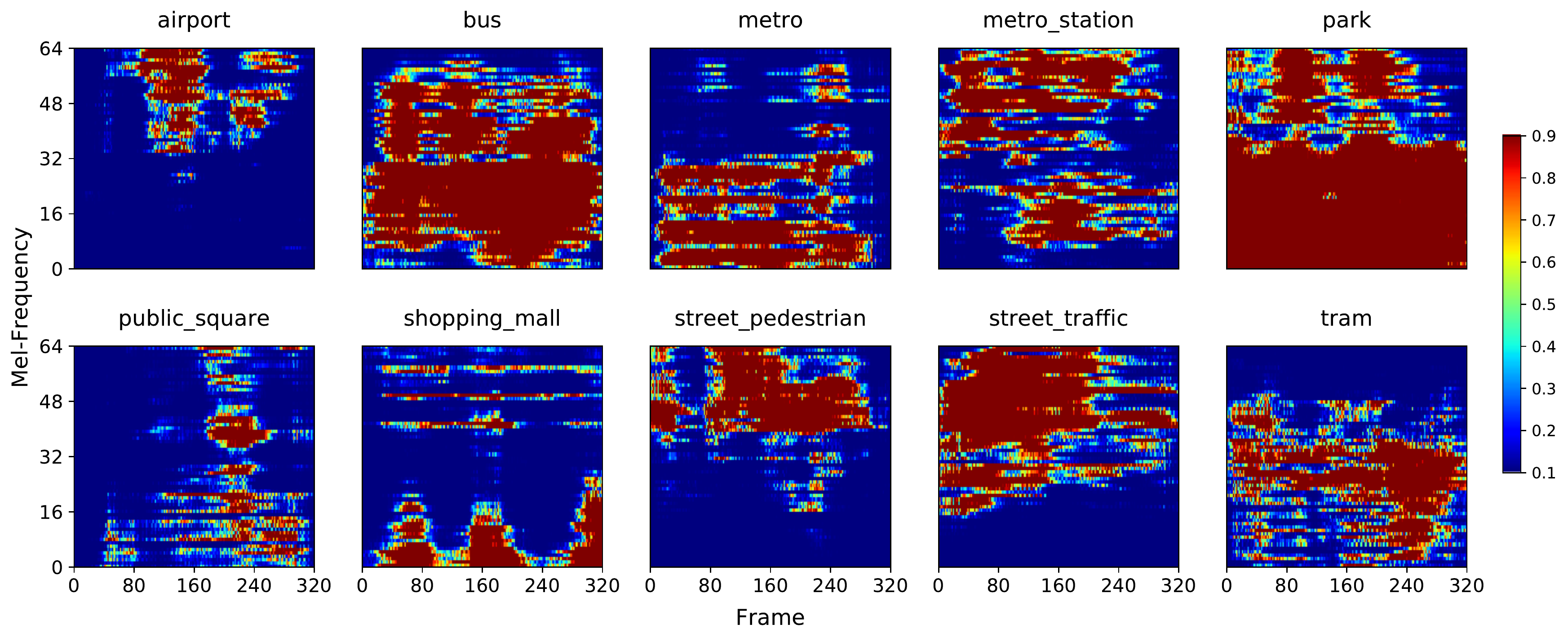}
    \centerline{(c) Heat maps of CNNs on data from device C}
  \end{minipage}
  \caption{Heat maps with a size of $64\times320$ are obtained by visualising the attention matrix $A$ in our multi-task CAA-Net which works on DCASE 2018 dataset. The horizontal and vertical axes represent the time frames and frequency bins, respectively. }
  \label{fig:heatmap}
\end{figure*}

%\begin{figure*}[!htbp]
%    \centering
%  \begin{subfigure}
%    \includegraphics[width=.99\linewidth]{IEEEtran/figures/heatmap-deviceA.pdf}
%    \subcaption{(a) Heat maps of CNNs on data from device A}
%  \end{subfigure}
%  \begin{subfigure}
%    \includegraphics[width=.99\linewidth]{IEEEtran/figures/heatmap-deviceB.pdf}
%    \subcaption{(b) Heat maps of CNNs on data from device B}
%  \end{subfigure}
%  \begin{subfigure}
%    \includegraphics[width=.99\linewidth]{IEEEtran/figures/heatmap-deviceC.pdf}
%    \subcaption{(c) Heat maps of CNNs on data from device C}
%  \end{subfigure}
%  \caption{Heat maps with a size of $64\times320$ are obtained by visualising the attention matrix $A$ in our multi-task CAA-Net which works on DCASE 2018 dataset. The horizontal and vertical axes represent the time frames and frequency bins, respectively. }
%  \label{fig:heatmap}
%\end{figure*}

%\subsection{Visualisation of Acoustic Scenes}
\subsection{Visualisation of CNNs}
\label{sec:visualise}

To visualise the internal convolutional layers of the CNNs, heat maps of the attention matrices in our proposed multi-task CAA-Net are shown in Fig.~\ref{fig:heatmap}. The heat maps are obtained from the learnt attention matrices in each class of acoustic scene recorded at the same time and place by the three devices. We can see that, the learnt attention matrices are mostly similar across the three devices. For some classes, such as \textit{airport}, \textit{metro}, \textit{public square}, and \textit{tram}, more time-frequency bins in the heat maps of device A have high values than in those of devices B and C. This may be caused by the fact that the data qualities differ from each other, as the recordings of the database were performed using three devices. Device A, as an in-ear device, has less noise than devices B and C, which both are mobile devices. The heat maps can reflect features of some special acoustic scenes. For instance, the heat maps of traffic scenes, such as \textit{bus}, \textit{metro}, and \textit{tram}, contain time-continuity bins, which is consistent with the characteristics of traffic scenes. In \textit{airport}, \textit{metro station}, \textit{park}, \textit{shopping mall}, \textit{street pedestrian}, and \textit{street traffic} classes, there are some non-traffic sounds like speech, bird sound, etc. Hence, the time-frequency bins in these scenes concentrate on the high-frequency areas. 

Visualisation of heat maps can provide a way to explain the decisions of CNNs in real-life applications, particularly in the security field. To ensure the machine is able to make an understandable and correct decision, it is essential to obtain an explanation of the decisions from the classifiers. Further, the visualisation technique offers the potential to improve the performance of other acoustic tasks such as in sound event detection, as the contribution weights of the heat maps might be related to the time and frequency of sound events.

\section{Conclusions and Future Work}
\label{sec:conclusion}
We proposed multi-task conditional atrous \textit{Convolutional Neural Networks} (CNNs) with attention, to visualise and understand acoustic scenes recorded with multiple devices. In the proposed model, log mel spectrograms were fed into two CNN models to predict device classes and acoustic scene classes. The predicted device classes were represented by one-hot encoders. The CNNs for acoustic scene classification were conditioned by feeding the transformation of device one-hot encoders into the convolutional layers. Four dilated convolutional layers, followed by a global attention pooling layer, were utilised to preserve the size of the feature maps for visualisation. Our proposed conditional training performed better than single-device training and joint training. Further, the intermediate layers of the CNNs were visualised and analysed in our work.

In future work, as the fixed weight of the loss function (the value of $\lambda$) is not friendly to optimise while training a neural network, and not flexible to adapt to a new dataset, we will explore learning the value of $\lambda$ during the training procedure, to obtain a more robust and flexible model able to work on different datasets. Due to the effect of type of scenes as aforementioned, a hierarchical structure to classify both the type of scenes and device classes will also be considered. Next, CNNs trained on one dataset which is collected with three concrete devices are challenging to be applied to datasets recorded by other devices. Therefore, training of cross-database CNN models will be considered, so that the data from more devices can be included. The computation complexity of our conditional training approach will be a burden if the data is recorded by many devices (more than three) in practice. In this regard, efficient training strategies such as transfer learning~\cite{yang2014cross}, and lifelong learning~\cite{qi2018life} are promising to achieve an efficient training procedure. We will also attempt to utilise the learnt heat maps for other acoustic tasks, such as audio event detection. 

% In future efforts, we will consider training cross-database CNN models so that the data from more devices are involved. Further, training strategies such as transfer learning~\cite{yang2014cross}, and life-long learning~\cite{qi2018life} will be considered to reduce the computation complexity of multi-device data. We will also attempt to utilise the learnt heat maps for other acoustic tasks, such as audio event detection.

\ifCLASSOPTIONcaptionsoff
  \newpage
\fi



\bibliographystyle{IEEEtran}
\bibliography{IEEEabrv,reference_arxiv}
%\begin{thebibliography}{1}


%\end{thebibliography}


%\clearpage
%\balance
\begin{IEEEbiography}[{\includegraphics[width=1in,height=1.25in,clip,keepaspectratio]{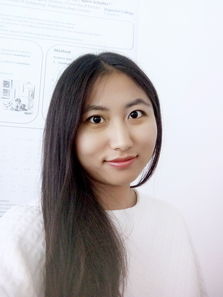}}]{Zhao Ren} (S'19) received her bachelor and master degrees in Computer Science and Technology from the Northwestern Polytechnical University (NWPU) in the P.R.\ China, in 2013 and 2017. Currently, she is an EU-researcher and working on her Ph.\,D.\ degree at the Chair of Embedded Intelligence for Health Care and Wellbeing, University of Augsburg, Germany, where she is involved with the H2020-MSCA-ITN-ETN project TAPAS, for pathological speech analysis. Her research interests mainly lie in developing deep learning techniques for the applications in health care and wellbeing. She regularly reviews IEEE Transactions on Cybernetics, IEEE Transactions on Neural Networks and Learning Systems, IEEE Transactions on Affective Computing, and IEEE Transactions on Multimedia, etc. She has (co-)authored publications in peer-reviewed journals and conference proceedings, including IEEE Journal of Biomedicaland Health Informatics, IEEE ICASSP, ISCA INTERSPEECH, DCASE, etc. 
\end{IEEEbiography}

\begin{IEEEbiography}[{\includegraphics[width=1in,height=1.25in,clip,keepaspectratio]{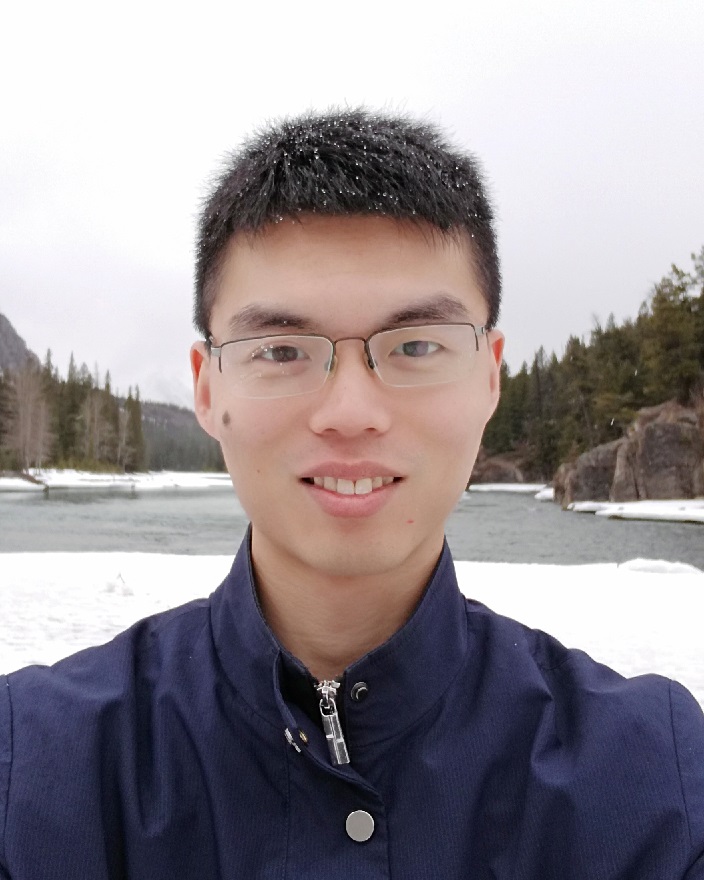}}] {Qiuqiang Kong} (S'17) received his Ph.D. degree from University of Surrey,
Guildford, UK in 2020. Following his PhD, he joined ByteDance AI Lab as a
research scientist. His research topic includes the classification,
detection and separation of general sounds and music. He is known for
developing attention neural networks for audio tagging, and winning the
audio tagging task in the detection and classification of acoustic scenes
and events (DCASE) challenge in 2017. He has co-authored paper in journals
and conferences including IEEE/ACM Transactions on Audio, Speech, and
Language Processing (TASLP), ICASSP, INTERSPEECH, IJCAI, DCASE, EUSIPCO,
LVA-ICA, etc. He was nominated as the postgraduate research student of the
year in University of Surrey, 2019. He is a frequent reviewer for over ten
world well known journals and conferences including TASLP, TMM, SPL, TKDD,
JASM, EURASIP, Neurocomputin, Neural Networks, ISMIR, CSMT, etc.
\end{IEEEbiography}

%\begin{IEEEbiography}[{\includegraphics[width=1in,height=1.25in,clip,keepaspectratio]{figures/kong.jpg}}]{Qiuqiang Kong} (S'17) received the B.Sc. and the M.E. degree in South China University of Technology, Guangzhou, China, in 2012 and 2015, respectively. He is currently pursuing a PhD degree at the University of Surrey, Guildford, UK. His research interest includes audio signal processing and neural networks.
%\end{IEEEbiography}

\begin{IEEEbiography}[{\includegraphics[width=1in,height=1.25in,clip,keepaspectratio]{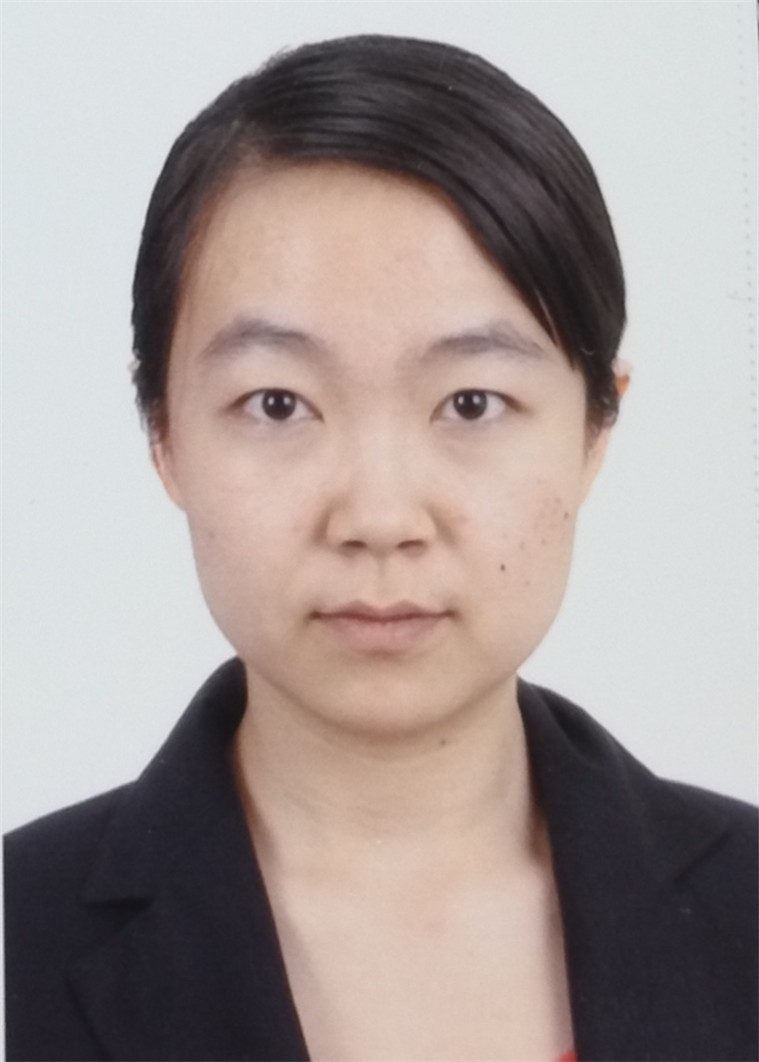}}]
{Jing Han} (S'16) is a postdoctoral researcher at the Department of Computer Science and Technology at the University of Cambridge, UK. She received her doctoral degree in Computer Science from the University of Augsburg, Germany, in 2020. Prior to that, she received her master degree from Nanyang Technological University, Singapore, in 2014, and her bachelor degree in electronic and information engineering from Harbin Engineering University, China, in 2011. Her research interests are related to deep learning for human-centric multimodal affective computing and health care. Besides, she co-chaired the 7th Audio/Visual Emotion Challenge (AVEC) and workshop in 2017, and served as a program committee member of AVEC challenge and workshop in 2018 and TPC member of ACM Multimedia in 2019 and 2020. She was awarded student travel grants from IEEE SPS and ISCA to attend ICASSP and INTERSPEECH in 2018. She (co-)authored more than 35 publications in peer-reviewed journals and conference proceedings, such as IEEE Transactions on Affective Computing, IEEE Transactions on Pattern Analysis and Machine Intelligence, IEEE Computational Intelligence Magazine, ACM Multimedia, IEEE ICASSP, ISCA INTERSPEECH.
%{\bf Jing Han} (S'16) received her bachelor degree (2011) in electronic and information engineering from Harbin Engineering University (HEU), China, and her master degree (2014) from Nanyang Technological University, Singapore. She is now working as a doctoral student with the Chair of Embedded Intelligence for Health Care and Wellbeing at the University of Augsburg, Germany, involved in two EU's Horizon 2020 projects SEWA and RADAR CNS. She reviews regularly for IEEE Transactions on Cybernetics, IEEE Access, and IEEE Signal Processing Letters. Her research interests are related to deep learning for multimodal affective computing and health care. Besides, she co-chaired the 7th Audio/Visual Emotion Challenge (AVEC) and workshop in 2017, and served as a program committee member of the 8th AVEC challenge and workshop in 2018. Moreover, she was awarded student travel grants from IEEE SPS and ISCA to attend ICASSP and INTERSPEECH in 2018.
\end{IEEEbiography}

%\begin{IEEEbiography}[{\includegraphics[width=1in,height=1.25in,clip,keepaspectratio]{figures/Mark_Plumbley.jpg}}]{Mark D. Plumbley}  (S'88-M'90-SM'12-F'15) received the B.A.(Hons.) degree in electrical sciences and the Ph.D. degree in neural networks from University of Cambridge, Cambridge, U.K., in 1984 and 1991, respectively. Following his PhD, he became a Lecturer at Kingís College London, before moving to Queen Mary University of London in 2002. He subsequently became Professor and Director of the Centre for Digital Music, before joining the University of Surrey in 2015 as Professor of Signal Processing. He is known for his work on analysis and processing of audio and music, using a wide range of signal processing techniques, including matrix factorization, sparse representations, and deep learning. He is a co-editor of the recent book on Computational Analysis of Sound Scenes and Events,
%and Co-Chair of the recent DCASE 2018 Workshop on Detection and Classifications of Acoustic Scenes and Events. He is a Member of the IEEE Signal Processing Society Technical Committee on Signal Processing Theory and Methods, and a Fellow of the IET and IEEE.
%\end{IEEEbiography}

\begin{IEEEbiography}[{\includegraphics[width=1in,height=1.25in,clip,keepaspectratio]{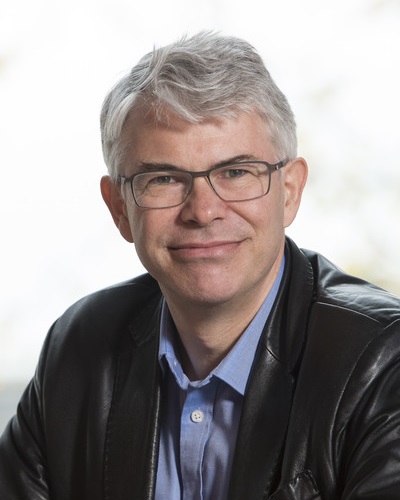}}]{Mark D. Plumbley} (S'88-M'90-SM'12-F'15) received
the B.A.(Hons.) degree in electrical sciences
and the Ph.D. degree in neural networks from University
of Cambridge, Cambridge, U.K., in 1984 and
1991, respectively. Following his PhD, he became a
Lecturer at Kingâ€™s College London, before moving
to Queen Mary University of London in 2002, where he
subsequently became Professor and Director of the
Centre for Digital Music. He joined the University
of Surrey in 2015 as Professor of Signal
Processing, and in 2019 became
Head of School of Computer Science and Electronic Engineering. 
He is known for his work on analysis
and processing of audio and music, using a wide range of signal processing
techniques, including matrix factorization, sparse representations, and deep
learning. He is a co-editor of a recent book on Computational Analysis of
Sound Scenes and Events, and Co-Chair of the recent DCASE 2018 Workshop
on Detection and Classifications of Acoustic Scenes and Events. He is a
Member of the IEEE Signal Processing Society Technical Committee on
Signal Processing Theory and Methods, and a Fellow of the IET and IEEE.
\end{IEEEbiography}

\begin{IEEEbiography}[{\includegraphics[width=1in,height=1.25in,clip,keepaspectratio]{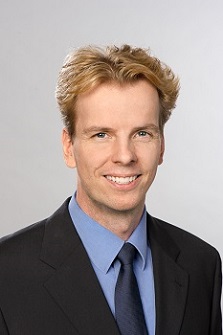}}]{Bj\"orn W. Schuller} (M'06--SM'15--F'18) received his diploma in 1999, his doctoral degree for his study on automatic speech and emotion recognition in 2006, and his habilitation and Adjunct Teaching Professorship in the subject area of signal processing and machine intelligence in 2012, all in electrical engineering and information technology from Technische Universit\"at M\"unchen (TUM), Germany. He is a tenured Full Professor heading the Chair of Embedded Intelligence for Health Care and Wellbeing, University of Augsburg, Germany, and a Professor of Artificial Intelligence heading GLAM -- the Group on Language, Audio \& Music, Department of Computing at the Imperial College London in London, UK. Dr.\,Schuller was an elected member of the IEEE Speech and Language Processing Technical Committee, Editor in Chief of the IEEE Transactions on Affective Computing, and is President-emeritus of the AAAC, Fellow of the IEEE, and Senior Member of the ACM. He (co-)authored 5 books and more than 800 publications in peer reviewed books, journals, and conference proceedings leading to more than 33,000 citations (h-index 79).
\end{IEEEbiography}

%\end{IEEEbiography}


%\newpage



%\vfill


\end{document}